\documentclass[twoside,twocolumn,9pt]{article}
\usepackage{extsizes}
\usepackage[super,sort&compress,comma]{natbib} 
\usepackage[version=3]{mhchem}
\usepackage[left=1.5cm, right=1.5cm, top=1.785cm, bottom=2.0cm]{geometry}
\usepackage{balance}
\usepackage{mathptmx}
\usepackage{sectsty}
\usepackage{graphicx} 
\usepackage{lastpage}
\usepackage[format=plain,justification=justified,singlelinecheck=false,font={stretch=1.125,small,sf},labelfont=bf,labelsep=space]{caption}
\usepackage{float}
\usepackage{fancyhdr}
\usepackage{fnpos}
\usepackage[english]{babel}
\addto{\captionsenglish}{%
  
}
\usepackage{array}
\usepackage{droidsans}
\usepackage{charter}
\usepackage[T1]{fontenc}
\usepackage[usenames,dvipsnames]{xcolor}
\usepackage{setspace}
\usepackage[compact]{titlesec}
\usepackage{hyperref}

\usepackage{epstopdf}

\usepackage[normalem]{ulem}
\usepackage{mathrsfs,bm,amssymb,amsmath}

\newcommand{\F}{\mathscr{F}}
\newcommand{\U}{\mathscr{U}}
\newcommand{\HH}{\mathscr{H}}
\newcommand\x{\bm{x}}
\newcommand\rme{\mathrm{e}}
\newcommand\be{\begin{equation}}
\newcommand\ee{\end{equation}}
\newcommand\rmd{\mathrm{d}}
\newcommand{\reff}[1]{(\ref{#1})}
\newcommand{\Cas}{{\rm Cas}}
\newcommand{\ccf}{CCF}
\newcommand{\bc}{BC}
\newcommand{\op}{OP}
\newcommand{\mc}{MC}
\newcommand{\vv}{\textsl{v}}
\usepackage{setspace} 
\begin{document}

\doublespacing

\pagestyle{fancy}
\thispagestyle{plain}
\fancypagestyle{plain}{
\renewcommand{\headrulewidth}{0pt}
}

\makeFNbottom
\makeatletter
\renewcommand\LARGE{\@setfontsize\LARGE{15pt}{17}}
\renewcommand\Large{\@setfontsize\Large{12pt}{14}}
\renewcommand\large{\@setfontsize\large{10pt}{12}}
\renewcommand\footnotesize{\@setfontsize\footnotesize{7pt}{10}}
\makeatother

\renewcommand{\thefootnote}{\fnsymbol{footnote}}
\renewcommand\footnoterule{\vspace*{1pt}%
\color{gray}\hrule width 3.5in height 0.4pt \color{black}\vspace*{5pt}} 
\setcounter{secnumdepth}{5}

\makeatletter 
\renewcommand\@biblabel[1]{#1}            
\renewcommand\@makefntext[1]%
{\noindent\makebox[0pt][r]{\@thefnmark\,}#1}
\makeatother 
\renewcommand{\figurename}{\small{Fig.}~}
\sectionfont{\sffamily\Large}
\subsectionfont{\normalsize}
\subsubsectionfont{\bf}
\setstretch{1.125} 
\setlength{\skip\footins}{0.8cm}
\setlength{\footnotesep}{0.25cm}
\setlength{\jot}{10pt}
\titlespacing*{\section}{0pt}{4pt}{4pt}
\titlespacing*{\subsection}{0pt}{15pt}{1pt}

\fancyfoot{}
\fancyfoot[RO]{\footnotesize{\sffamily{\thepage}}}
\fancyfoot[LE]{\footnotesize{\sffamily{\thepage}}}
\fancyhead{}
\renewcommand{\headrulewidth}{0pt} 
\renewcommand{\footrulewidth}{0pt}
\setlength{\arrayrulewidth}{1pt}
\setlength{\columnsep}{6.5mm}
\setlength\bibsep{1pt}

\twocolumn[
  \begin{@twocolumnfalse}{}
\vspace{1em}
\sffamily
\begin{tabular}{p{18cm}}
\LARGE{\textbf{Critical Casimir forces in soft matter}} \\
\vspace{0.3cm} \\
\large{A. Gambassi\textit{$^{a}$} and S. Dietrich\textit{$^{b}$}} \\
\vspace{0.3cm} \\
\normalsize{We review recent advances in the theoretical, numerical, and experimental studies of critical Casimir forces in soft matter, with particular emphasis on their relevance for the structures of colloidal suspensions and on their dynamics. Distinct from other interactions which act in soft matter, such as electrostatic and van der Waals forces, critical Casimir forces are effective interactions characterised by the possibility to control reversibly their strength via minute temperature changes, while their attractive or repulsive character is conveniently determined via surface treatments or by structuring the involved surfaces. These features make critical Casimir forces excellent candidates for controlling the equilibrium and dynamical properties of individual colloids or colloidal dispersions as well as for possible applications in micro-mechanical systems. In the past 25 years a number of theoretical and experimental studies have been devoted to investigate these forces primarily under thermal equilibrium conditions, while their dynamical and non-equilibrium behaviour is a largely unexplored subject open for future investigations.} 
\end{tabular}

 \end{@twocolumnfalse} \vspace{0.6cm}

  ]

\renewcommand*\rmdefault{bch}\normalfont\upshape
\rmfamily
\section*{}
\vspace{-1cm}

\footnotetext{\textit{$^{a}$~SISSA --- International School for Advanced Studies and INFN,
via Bonomea 265, 34136 Trieste, Italy. E-mail: gambassi@sissa.it}}
\footnotetext{\textit{$^{b}$~Max-Planck Institute for Intelligent Systems, Heisenbergstr.~3, 70569 Stuttgart, Germany and
Institut f\"ur Theoretische Physik IV,  Universit\"at Stuttgart, Pfaffenwaldring 57, 70569 Stuttgart, Germany. E-mail: dietrich@is.mpg.de}}

\tableofcontents


\section{Introduction}
\label{sec:intro}

Critical Casimir forces (CCFs) emerge at the surfaces of objects immersed in a fluid medium whenever they confine the fluctuations of the order parameter (OP) of a continuous phase transition occurring within that medium.
Accordingly, these forces are deeply related to phase transitions and critical phenomena in statistical systems due to the effects of spatial confinements on them\cite{FB72}, i.e., two topics which are part of the scientific legacy of 
Michael E.~Fisher.
This is the reason why we deem it natural to dedicate this review to his memory.

In fact, the very same idea of the emergence of peculiar solvation forces in a critical medium --- which, later on, have been called \emph{critical Casimir forces} --- can be traced back to a seminal paper\cite{FG78} written in 1978 by Fisher and Pierre-Gilles de Gennes. 
These forces owe their name to the fact that they are the thermodynamicly analogous ones related to the celebrated Casimir effect of quantum electrodynamics\cite{C48}, in which the relevant fluctuations are those of quantum nature of the electromagnetic fields in vacuum, while the confining surfaces are provided by perfectly conducting metallic plates. 
As in the latter case, the dependence of {\ccf}s on the relevant physical constants and geometrical parameters characterizing the system is largely universal, in a sense we shall specify below. 

The magnitude of {\ccf}s can be estimated on the basis of dimensional analysis: as they are essentially determined by thermal fluctuations of an {\op}, their typical energy scale is set by the thermal energy $k_BT$ at temperature $T$. 
In turn, this is the scale typically involved in the physics of mesoscopic particles such as \emph{colloids}, having  a radius in the micrometer and sub-micrometer range.
Therefore {\ccf}s are 
expected to give rise to relevant effects in soft matter, which primarily involves colloids dispersed in a liquid medium. 
For this reason the present contribution focuses on {\ccf}s in soft matter, with particular emphasis on colloidal suspensions. 
However, we shall also discuss yet another important manifestation of {\ccf}s on the behaviour of fluids at the microscopic scale, i.e., how they affect  the thickness of complete wetting films.
From the investigations reviewed here, it clearly emerges that, different from the typical interactions in soft matter\cite{I2011,P2006}, {\ccf}s offer a large degree of tunability which is already finding interesting practical applications. 

This review is organised as follows: In Sec.~\ref{sec:FSS-C} we briefly recall 
(following Ref.\!\nocite{GD2011}\citenum{GD2011})
 the basic notions of statistical physics of spatially confined systems and discuss the effects of introducing physical boundaries on their behaviour, explaining if and how {\ccf}s emerge. As anticipated, these forces are characterised by universal scaling functions, the analytical and numerical determination of which are presented in Sec.~\ref{sec:SF}.  
While most of these predictions refer to 
critical media confined to the film geometry,  in Sec.~\ref{sec:FtoC} we 
discuss how this knowledge can be used to study the case
in which curved or inhomogeneous surfaces are involved,
e.g., spherical and non-spherical colloids or patterned substrates of experimental relevance. 
Increasing the complexity of the system  further, we consider many-body effects of these forces as well as their subtle interplay with other forces which are usually at play in soft matter, such as electrostatic interactions in the presence of electrolytes. In Sec.~\ref{sec:exp} we review how the critical Casimir effect has been detected experimentally,  starting from what is arguably its earliest evidence\cite{BE85} dating back to 1985, to the first indirect, quantitatively reliable  measurement\cite{GC99} in 1999, 
and finally to the direct measurement\cite{HHGDB2008} in 2008. We provide also an overview of the experimental investigations of {\ccf}s in the presence of patterned substrates as well as in more complex solvents, i.e., if in addition van der Waals and electrostatic interactions are present. Following this overview of the equilibrium features of {\ccf}s, we review our current theoretical, numerical, and experimental understanding of their dynamical behaviour which, beyond the cases of specific aspects or toy models, is still a largely unexplored subject. In Sec.~\ref{sec:persp} we provide our perspective on this subject, mentioning some possible further applications of {\ccf}s in soft matter. 

A general introduction to {\ccf}s can be found in the early Refs.\!\nocite{K94}\citenum{K94}, 
\!\!\nocite{BDT2000}\citenum{BDT2000}
and in the subsequent short reviews in 
Refs.\!\nocite{G2009}\citenum{G2009} and \!\nocite{GD2011}\citenum{GD2011}.
More recently, some specific aspects of {\ccf}s have been reviewed, including their role in determining the phase behaviour of colloidal suspensions,\cite{MD2018} establishing exact analytical results,\cite{DD2022} and measurements involving optical trapping.\cite{CMGV-2021} Accordingly, our focus here is primarily on aspects which have not yet been covered in the above references. 
In order to keep the presentation focussed on the most relevant aspects of {\ccf}s,  the available literature will not be reviewed completely, but pertinent studies will be at least quoted, as guidance for the interested reader.


\section{Statistical systems under confinement and the critical Casimir effect}
\label{sec:FSS-C}

The physical properties of soft matter are determined by various effective interactions, which depend on its structure at various length scales and on its thermodynamic conditions. For example, in colloidal suspensions, a prominent role is played by screened electrostatic, hydrodynamic, and depletion forces. 
As anticipated in Sec.~\ref{sec:intro}, thermal fluctuations can also generate effective forces\cite{FG78}, in analogy to the celebrated Casimir effect in quantum electrodynamics\cite{C48}.
In particular, the latter effective forces are heuristically expected to become particularly relevant whenever the thermal fluctuations are pronounced and spatially correlated, in such a way that they influence both qualitatively and quantitatively the thermodynamic and structural properties of the fluctuating medium. This is known to be the case upon approaching a continuous (i.e., second-order) phase transition of the medium, close to which the relevant thermal fluctuations are those of a spatially varying \op\  $\phi(\x)$ associated with the phase transition\cite{G2009}.

\subsection{Critical phenomena and universality} 
\label{sec:CP-U}

Upon approaching the critical point of a continuous phase transition, the thermal fluctuations of the order
parameter $\phi(\x)$ become spatially correlated across a typical distance $\xi$. Due to the emergence of a collective behaviour, this correlation length actually diverges algebraically upon approaching the phase transition. 
Accordingly, close to a critical point, $\xi$ eventually becomes the dominant length scale. It is significantly larger than the microscopic scales characterising the molecular structure of the medium (and, possibly, larger than the scales of other effective interactions at work in the system). 
In this case, the free energy $\F$ of the medium itself receives a relevant non-analytic contribution due to the thermal fluctuations of
 $\phi(\x)$. The physical nature of the quantity playing the role of the {\op} depends on the specific critical point of interest: In the case of the normal-to-superfluid transition of liquid ${}^4$He, $\phi$ is identified with the quantum wave-function of the superfluid, whereas for the demixing transition of a classical binary liquid mixture --- which we primarily consider here, --- $\phi$ is proportional to the difference $c_A(\x)-c_B(\x)$
between the local concentrations $c_{A,B}$ of the two species $A$ and $B$ forming the
mixture.
The thermal fluctuations of  $\phi$ are controlled by an effective Hamiltonian $\HH[\phi]$  which determines their probability $\propto \rme^{-\HH[\phi]}$ and thus the effective free energy  
$\F = - k_BT\ln \int [\rmd\phi] {\rm e}^{-\HH[\phi]}$ of the medium.
In principle, the effective Hamiltonian $\HH[\phi]$ can be derived by coarse-graining the microscopic Hamiltonian of the system. Therefore it is expected to depend on the structure of the medium at various length scales and on its 
specific material and microscopic
properties. 
However, this generally prohibitive task turns out to be unnecessary upon approaching the critical point because 
only few of these features are relevant and thus eventually crucial in determining the thermodynamic behaviour at mesoscopic length
scales. 
Accordingly, the form of $\HH[\phi]$ can be drastically simplified and only its dependence on a few gross features of the medium, such as the ranges and the symmetries of the microscopic interactions, and on general properties of
the order parameter, e.g., the number of its components and the nature of the broken
symmetry, matter.
After this reduction, statistical systems, which differ significantly at the
microscopic level, eventually display a common singular behaviour upon approaching a
critical point --- a fact encoded in the notion of \emph{universality}
and universality classes within critical phenomena.
This means, for example, that upon approaching the temperature $T_c$ of the critical
point from the homogeneous, i.e., disordered phase, the correlation length $\xi$ as a function of
temperature $T$ increases algebraically as $\xi \simeq
\xi_0^+|T/T_c-1|^{-\nu}$, with a critical exponent $\nu$ and a typical molecular scale $\xi_0^+$ of the medium. Whereas $T_c$ and $\xi_0^+$ depend on the material properties of the medium undergoing the
transition, the exponent $\nu$ does not and is therefore universal. For example,
$\nu$ takes the same value for all binary liquid mixtures undergoing a second-order
demixing transition and  therefore it is universal.

If an object, such as a plate or a particle, is immersed into the fluid medium, the {\op}  and its thermal fluctuations are affected by the presence of the additional interaction between the molecules of the object and those of the medium. In particular, the resulting perturbation is expected to extend into the medium by a typical distance around the object of the order of the correlation length $\xi$. 
As in the case of bulk properties, it turns out that, 
upon approaching the critical point, 
the microscopic details of the interaction between the object and the medium are largely irrelevant for determining the behaviour  of this inhomogeneous system at mesoscopic scales. This observation leads naturally to an extension of the notion of universality to the case of surface critical behaviour and therefore to the introduction of the so-called
\emph{surface universality classes}\cite{Bb83,D86}. 
In fact, for the practical purposes of a coarse-grained description, the presence of such objects results into suitable effective boundary conditions ({\bc}s) for the {\op}  $\phi$ at their surfaces.
In particular, the interaction with the object can induce a local enhancement of $\phi$ upon approaching its surface, such that close to it either positive or negative values of $\phi$ are favored. These adsorption preferences effectively lead to the {\bc} which is usually denoted by $(+)$ or $(-)$, respectively. Alternatively, there might be no preferential adsorption, leading to the Dirichlet $(O)$ {\bc} $\phi=0$ at the surface. 
In the presence of enhanced interactions among the ordering degrees of freedom near the surface one can also encounter effective Neumann {\bc}s which, beyond mean-field theory, eventually lead to a multicritical, so-called surface-bulk (SB) surface transition\cite{Bb83,D86,D2020}. 
The generic inequality of the interactions between the molecules of the confining solid wall and those two types of molecules forming the two components of a classical binary liquid mixture results in the $(+)$ or the $(-)$ {\bc}. Concerning quantum fluids, the necessary continuity of the 
wave-function causes generically  $\phi$ (e.g., for the superfluid transition in $^4$He) to vanish at the surface of the immersed object. However, exceptions may emerge\cite{MD2006}  upon confining quantum mixtures such as those of 
$^3$He and $^4$He.

\subsection{Spatial confinement and the critical Casimir effect} 
\label{subsec:CE}

If a second object is immersed into the fluid medium, it interacts locally with  $\phi$ and its fluctuations, which are possibly affected by the presence of the first object. 
In particular, this is expected to be the case whenever the distance $L$ of closest approach between the two objects is comparable to or smaller than the correlation length $\xi$,  such that the regions of the medium, within which each surface separately affects its surrounding, overlap. If, on the contrary,  $L\gg \xi$, the second object interacts with a medium which is not significantly perturbed by the first one, yielding a negligible effective interaction between the objects, the range of which is therefore set by $\xi$.
Similarly, immersing a third object, sufficiently close to the previous two, will also affect their mutual interaction, leading to non-additive many-body effects, which will be discussed in Sec.~\ref{ssec:FtoC-mb}. 
This effective interaction, mediated by the {\op}  and its near-critical fluctuations --- known as the \emph{critical Casimir effect}, --- inherits the universality of the critical behaviours in the bulk and at the surfaces. Accordingly, the associated force depends only on the bulk and surface universality classes, the latter being characterised by the effective {\bc} imposed on the {\op}  at the surfaces, as well as on the shape of the boundaries (see Sec.~\ref{sec:FtoC}).
The very thermodynamic nature of the critical Casimir effect can be understood by considering the grand canonical free energy $\Omega_V$ of a fluid medium in thermal equilibrium at temperature $T$, which, in the film geometry,  is confined by two parallel surfaces of large area $A$, separated by a distance $L$, and with total volume $V = A L$. Upon increasing $L$,  the free energy  naturally decomposes as
\be
\frac{\Omega_V(L,T)}{A} = - L p_b(T) +  \Omega_s(T) + \frac{k_{\rm B}T}{L^2}X_{\rm
ex}(L/\xi) + \mbox{corrections},
\label{eq:Fdec}
\ee
where we assume $L,\xi\gg\xi_0^+$, and $A\rightarrow\infty$. 
Here we focus on the non-analytic contribution $\Omega_V$ to the actual free energy of the system, which contains also a non-singular background term due to the non-critical degrees of freedom and fluctuations within the system. The latter typically result into other effective interactions (usually referred to as \emph{solvation forces}) which also emerge as described below. 
Referring to $\Omega_V$, the fact that the dependence of $X_{\rm ex}$ on $T$ occurs only via the correlation length $\xi(T)$ is actually a consequence of finite-size scaling in critical phenomena\cite{FB72,Br82,PF83,C88,BDT2000}. 
The first term on the r.h.s.~of Eq.~\reff{eq:Fdec} corresponds to the free energy $-A L p_b$  which a volume $V=AL$ of the
medium contained inside the film  would have in the bulk, i.e., in the absence of confining surfaces; accordingly, $p_b$ is the bulk pressure. The second term corresponds to the additional total surface free energy $A \Omega_s(T)$, with $\Omega_s(T) = \Omega_{s,1}(T)+\Omega_{s,2}(T)$ due to the independent (i.e., for $L\rightarrow\infty$) 
presence of the two opposing and confining surfaces within the medium, facing each other and contributing with $\Omega_{s,1}$ and $\Omega_{s,2}$.
The third term corresponds to the interaction free
energy (also referred to as excess free energy) due to the simultaneous presence of the two plates, the temperature
dependence of which is encoded solely in the dimensionless, \emph{universal} scaling function
$X_{\rm ex}$, along with its dependence on the effective {\bc}.
To be concrete, here and in the following we consider the experimentally relevant case of statistical systems in spatial dimension $d=3$, although {\ccf}s have been investigated in various spatial dimensions, especially within the field-theoretical context.
As a consequence of the dependence of $\Omega_V$ on $L$ and of the decomposition in Eq.~\reff{eq:Fdec}, 
each of the two
confining surfaces is subject to a force
\be
\label{first_appearance}
F_V = - \partial \Omega_V/\partial L = A
p_b(T) + F_\Cas+ \mbox{corrections},
\ee
pointing from inside the film towards its outside, where $F_\Cas$ 
is the {\ccf}
\begin{equation}
F_\Cas  = -  A \, \frac{\partial}{\partial L} \left[\frac{k_{\rm
B}T}{L^2} X_{\rm ex}(L/\xi)\right] \equiv A\frac{k_{\rm
B}T}{L^3}X_{\Cas}(L/\xi),
\label{eq:FC}
\end{equation}
which vanishes as $L\to\infty$. 
Consider now the case 
of a confining surface of large but finite lateral extension, which
is completely surrounded by the fluctuating medium. The force $F_{\rm in}$ exerted on this wall from inside the film towards its outside is given by $F_{\rm in} = F_V(L)$, while the force $F_{\rm out}$ from outside towards the film inside and due to the unconfined medium outside the film is given by $F_{\rm out} = F_V(L\rightarrow\infty)$. Accordingly, using also Eq.~\eqref{first_appearance}, the \emph{total} force acting on the wall is $F_{\rm in} - F_{\rm out} = F_\Cas+ \mbox{corrections}$, i.e., its leading term as $L$ increases is actually provided by the {\ccf}. 
The analogies and differences between the expression in Eq.~\eqref{eq:FC} for the {\ccf}  and the corresponding one for the Casimir effect in quantum electrodynamics are discussed, e.g., in Ref.\!\nocite{G2009}\citenum{G2009}. 
As anticipated, the range of $F_\Cas$ is set by the correlation length $\xi$. The scaling function $X_{\rm Cas}$ is universal in the sense that it depends only on the gross features of the critical
behaviour of the system (i.e., the bulk universality class) and on the kind of effective {\bc}s imposed by the two plates on the {\op}  $\phi$, i.e., $(+,+)$ or $(+,-)$ for a binary liquid mixture of
classical fluids and $(O,O)$ for pure $^4$He.
Analogously to the cases of bulk systems and surfaces, the emerging universality allows one to predict the behaviour of the scaling function upon approaching a critical point in terms of suitably simplified models, which act as representatives of the corresponding universality class and thus retain all the relevant gross features of the physical system under study.
In fact, numerical simulations of lattice models and analytic studies of continuum theories for the {\op}  field $\phi$ both in the bulk\cite{P88,ZJ2002,LB2005} and at surfaces\cite{D86}, have provided significant insight
into the critical behaviour of a variety of different physical systems and yield
remarkably accurate theoretical predictions both for exponents and scaling
functions (see, e.g., Ref.\!\nocite{PV2002}\citenum{PV2002}).  
For example, the bulk
critical properties of a demixing classical binary  liquid mixture (as well as those of ferromagnetic/paramagnetic transitions) are captured on a lattice by the Ising
model (with ${\mathbb Z}_2$ symmetry) of one-component spins, whereas those of
the normal-superfluid transition of $^4$He are captured by the XY model (with $O(2)$ symmetry)  of
two-component spins. 
Alternatively, these two cases can be studied via the Landau-Ginzburg-Wilson effective Hamiltonian $\HH[\phi]$ of an $n$-component field $\phi$ with a $(\phi^2)^2$ self-interaction, $O(n)$ symmetry, for $n=1$ and $n=2$, respectively. 
In Sec.~\ref{sec:SF} below we shall discuss how universality has been exploited, via analytical and numerical techniques, in order  to determine the scaling function $X_\Cas$  in Eq.~\eqref{eq:FC}.
%


The universal scaling functions $X_{\rm ex}$ and $X_\Cas$ in Eqs.~\eqref{eq:Fdec} and \eqref{eq:FC}, respectively, depend on the scaling variable $y=L/\xi$. As a function of the temperature $T$, $y$ may assume the same value at two temperatures, one above and one below $T_c$, which correspond to the same correlation length $\xi$. In order to distinguish these two cases, sometimes the equivalent scaling variable $x = (T/T_c-1)(L/\xi_0^+)^{1/\nu}$ is considered. 

Before discussing in the next sections the characteristic features of {\ccf}s, we emphasise that these forces are fluctuating quantities themselves  because, as explained above, they arise from confining the thermal fluctuations of an order parameter.
While their mean value, given in Eq.~\eqref{eq:FC}, is characterised by a large degree of universality, their variance depends on the microscopic details of the system and is generically large compared to the average\cite{BAFG2002}. This raises the question of how this fact affects the numerical and experimental determination of the scaling function $X_\Cas$, which is addressed in Ref.\!\nocite{GGD21}\citenum{GGD21}.

\section{Scaling function: renormalisation group and numerical studies}
\label{sec:SF}

Similarly to what occurs for critical phenomena in the bulk, universality allows one to determine the scaling function $X_{\rm Cas}$ of the {\ccf}  in the cases mentioned above by studying the most convenient representative model belonging to the universality class under investigation. 
In particular, as a first step, one has to consider this model for the slab geometry, with macroscopically large transverse area $A$, finite thickness $L$, and with {\bc}s which realisrealisee the surface universality classes of interest.  
For example,  upon approaching the demixing transition of a classical binary liquid mixture, $X_{\rm Cas}$ can be determined by studying the Ising model (or any other spin model within the same universality class) in the slab geometry with fixed spins at the boundaries, pointing either in the same direction or in the opposite one, such as to realise effectively $(+,+)$ or $(+,-)$ 
{\bc}s. 
5
Equivalently, this model can be investigated  by considering the Landau-Ginzburg-Wilson effective Hamiltonian $\HH[\phi]$ for a scalar {\op}  $\phi$, within a slab with suitable surface fields at the boundaries.  
Similarly, $X_{\rm Cas}$  for the  superfluid transition in $^4$He can be determined by studying the XY model in a slab  with unconstrained spins at the boundaries, which eventually realise $(O,O)$ {\bc}.  Equivalently, this problem can be investigated  by considering $\HH[\phi]$ for a two-component order parameter, within a slab geometry, and with Dirichlet {\bc}s.  %

In view of the illustration in Sec.~\ref{sec:exp} of some experimental evidences for the occurrence of {\ccf}s, the previous discussion primarily focused on a medium undergoing a continuous phase transition which belongs to the universality class of either the normal-superfluid transition in $^4$He (XY universality class) or the demixing transition of classical binary liquid mixtures (Ising universality class).  In particular, we considered only those effective {\bc}s, i.e., surface universality classes,  which generically characterise the possible experimental realizations of these classes. As a rule of thumb, here we anticipate that the {\ccf} turns out to be attractive for equal {\bc}s [$(+,+)$, $(O,O)$] and repulsive for unequal {\bc}s [$(+,-)$], independently of the actual bulk universality class of the phase transition under investigation. 

In the present section we review the available knowledge about the scaling function $X_{\rm Cas}$ of the {\ccf}. In particular, in Subsec.~\ref{ssec:SF-ana} we consider various analytical studies --- primarily based on a field-theoretical approach, while a number of exact predictions based on lattice models in two bulk spatial dimensions ($d=2$) or based on exactly solvable models in general $d$ have been reviewed recently in Ref.\!\nocite{DD2022}\citenum{DD2022}.
In Subsec.~\ref{ssec:SF-num}, instead, we focus on the predictions which were obtained on the basis of Monte Carlo ({\mc})  simulations of lattice models and which provide quantitatively accurate estimates for $X_{\rm Cas}$, which are generically out of reach of analytical approaches, especially in the experimentally relevant case $d=3$.

\subsection{Analytical approaches}
\label{ssec:SF-ana}

Early theoretical investigations of the critical Casimir effect in the film geometry
were based primarily on field-theoretical approaches\cite{S81,INW86,KD92b}
which build upon previous analyses based on the minimization of a
Landau-Ginzburg-Wilson functional for $\HH[\phi]$\cite{FG78,INW86}. 
In particular, these studies can be cast into the general framework in which one considers a field theory in a bounded domain, in the presence of contributions localised at the boundaries. As in the bulk, and in view of universality, the form of these terms can be deduced from symmetry arguments and from power counting. For the simple but important case of a scalar {\op}  $\phi$, one finds that the effective free energy to be considered is given by
\be
\begin{split}
\HH[\phi] &= \int_{\mathscr{V}}\rmd^d x\, \left\{ \frac{1}{2} (\nabla \phi)^2 + \frac{r}{2}\phi^2 + \frac{u}{4!} \left(\phi^2\right)^2\right\} \\
&\quad\quad+ \sum_{i=1}^2\int_{\mathscr{A}_i}\rmd^{d-1} x_\| \left\{\frac{c_i}{2} \phi^2 - h_i \phi\right\},
\end{split}
\label{eq:LGWwB}
\ee
where the confined volume corresponds to $\mathscr{V} = [0,L]\times\mathscr{A}$ with $|\mathscr{A}| = |\mathscr{A}_1| = |\mathscr{A}_2| = A$ and with the same shape as $\mathscr{A}_i$, while the boundary contributions to the action are collected in the second term in which $\x = (x_\perp,\x_\|)$ with $x_\perp = 0$ for $\x_\| \in \mathscr{A}_1$ and $x_\perp = L$ for $\x_\| \in \mathscr{A}_2$.
In addition to the usual couplings in the bulk, which are controlled by the parameters $r$ (proportional to the actual reduced  temperature $(T-T_c)/T_c$ of the system) and $u>0$, the so-called \emph{surface enhancements} $c_{1,2}$ and the surface fields $h_{1,2}$ appear at the boundaries. Eventually they determine the behaviour of  $\phi(\x)$ upon 
approaching the boundaries, i.e., for $x_\perp \to 0$ or $x_\perp \to L$, and thus they generate the effective {\bc}s (see Refs.\!\nocite{D86}\citenum{D86}
and 
\!\nocite{D1997}\citenum{D1997}).
In particular, in the absence of symmetry-breaking terms at the boundaries, i.e., for $h_i = 0$, the renormalization-group fixed-point values $c^*_i$ of interest of the coupling $c_i$ turns out to be $c^*_i=+\infty$. This results (as anticipated above) into Dirichlet {\bc}s for the (fluctuating) field $\phi(\x)$ such that the mean value $\langle \phi(\x)\rangle$ vanishes upon approaching the boundary with a universal behaviour, which requires the introduction of a novel critical \emph{surface} exponent $\beta_1$ in addition to the bulk ones.
If, instead, $h_i \neq 0$, the symmetry breaking at the boundary implies, at bulk criticality and at the renormalization-group fixed point, an enhancement of $\langle \phi(\x)\rangle$ of the form $\langle \phi(\x)\rangle \sim x_\perp^{-\beta/\nu}$  upon approaching the boundary, where here $x_\perp$ is the spatial distance from the nearest boundary. The sign of $h_i$ equals the sign of the {\op} enhancement near the surface and thus renders the $(+)$ or the $(-)$ effective {\bc}. 
Depending on the choices made for $c_{1,2}$ and $h_{1,2}$, it is possible to realise various combinations of effective 
{\bc}s at the two surfaces of the film. 

As mentioned above, the functional in Eq.~\eqref{eq:LGWwB} and generalizations thereof (e.g., accounting for an $n$-component order parameter, which requires some care in specifying the boundary terms) form the basis for a number of theoretical investigations of {\ccf}s, of either analytic or numerical character (see below). 

\subsubsection{Mean-field approximation}
\label{sssec-MFT}
Concerning the analysis of the free energy $\Omega_V$  of a system, described by $\HH[\phi]$ and confined to a finite volume $V$, the simplest corresponding approximation consists of neglecting the statistical fluctuations in calculating the partition function $\int[\rmd \phi] \rme^{- \HH[\phi]}$. This implies that the free energy $\Omega_V$ is then given by $\Omega_V = \HH[\bar\phi]$, where 
$\bar \phi(\x)$ is the $L$-dependent configuration which minimises $\HH[\phi]$ and which satisfies the {\bc}s at the confining walls $\mathscr{A}_{1,2}$.
The dependence on $L$ of this mean-field free energy $\Omega_V$ of the confined system provides access to the mean-field {\ccf}  and the corresponding scaling function $X_{\rm ex}(L/\xi)$ according to Eqs.~\eqref{eq:Fdec} and \eqref{first_appearance}, in which --- consistent with the mean-field approximation --- one has $\xi \propto |r|^{-1/2}$. 
This approach requires the knowledge of the {\op}  configuration $\bar\phi(\x)$ in the whole film in order to be able to determine the critical Casimir pressure on the confining walls, i.e., on a well-localised part. In certain cases (involving both analytical and numerical calculations), this may constitute a serious drawback which can be circumvented by resorting to the notion of the stress tensor.
In fact, the stress tensor $T_{ij}(\x)$ (with $i$ and $j$ running over the $d$ components of the position vector) is a suitable space-dependent quantity which is constructed on the basis of the (fluctuating) {\op}  field $\phi(\x)$ and its spatial derivatives, such that the expectation value of its components renders the pressure locally exerted at position $\x$ on the boundary.
In fact, the corresponding force $\rmd{\bf f}$ acting on a small surface element with lateral extension $\rmd A$, centred at point $\x$, and belonging to the boundary of the system, is given by 
\be
\rmd f_i = \rmd A \sum_{j=1}^d \langle T_{ij} (\x)\rangle \hat n_j,
\label{eq:ST}
\ee
where ${\bf \hat n}$ is the unit normal vector pointing towards the portion of space occupied by the field.
Accordingly, in order to obtain the pressure locally exerted around a certain point at the boundary, 
it is sufficient to know $\phi(\x)$ (and its fluctuations) right at that boundary. In addition, under thermal equilibrium conditions, mechanical equilibrium within the system requires that the stress tensor is conserved and therefore $\langle \partial_j T_{ij}\rangle = 0$. (This follows also formally  from the fact that $T_{ij}$ can be viewed as a conserved Noether current 
associated with the lateral translational invariance.)
The explicit expression of $T_{ij}$ in terms of the (fluctuating) {\op}  field $\phi(\x)$ follows from Noether's theorem\cite{K94}. %
Within mean-field theory it is given by $T_{ij} = [\delta \HH/\delta (\partial_i\phi)] \partial_j\phi - \delta_{i,j} \HH $ in terms of $\HH$.
With this, the expectation value in Eq.~\eqref{eq:ST} can be calculated as for any local quantity. In particular, within the mean-field approximation discussed here, $\langle T_{ij}\rangle$ coincides with the value which $T_{ij}$ assumes for the given  mean-field  profile $\bar \phi(\x)$. The conservation of 
$T_{ij}$ actually allows one to calculate the pressure on the boundaries of the film by considering the value which $\bar \phi(\x)$ has on a suitably displaced boundary. This fact is useful especially for numerical calculations as well as analytic ones, whenever $\bar \phi(\x)$ is singular upon approaching the boundaries (as in the case of $(\pm)$ {\bc}s).
In particular, the determination of the {\ccf} for films via $T_{ij}$ can be easily generalised to  more complex geometries and topographycally or chemically inhomogeneous surfaces, which will be discussed in Sec.~\ref{ssec:FtoC-patter}. 

The expression of  $X_{\rm ex}(L/\xi)$, derived as discussed above within the mean-field approximation, generically depends on the values of the bulk coupling constant $u$ and the surface couplings $(c_i,h_i)$, which appear in Eq.~\eqref{eq:LGWwB} and which are left undetermined within the mean-field approximation. 
Accordingly, in order to derive a quantitative estimate of $X_{\rm ex}$ to be compared with 
{\mc} simulations or experimental data, two strategies are possible: 
(i)  improve the mean-field estimate by tuning the various constants to their renormalisation-group fixed-point values, which are independently known from the studies of bulk and surface criticality 
(see, e.g., Refs.~\citenum{PV2002} and~\citenum{D1997}); 
however, this approach generically renders inaccurate values. Alternatively, one can (ii) normalise  the mean-field prediction for  $X_{\rm ex}(L/\xi)$ with the critical Casimir amplitude $X_{\rm ex}(0)$ and use this \emph{ratio} in order to estimate the scaling function reduced this way.

Among the available mean-field predictions (see Ref.~\citenum{DD2022}
for a comprehensive review) obtained as described above, we briefly discuss here those which turned out to be relevant for a qualitative understanding of the {\ccf}  in experimentally accessible systems, such as critical binary liquid mixtures and superfluids.

In particular, in Ref.~\citenum{K97}
the effective Hamiltonian in Eq.~\eqref{eq:LGWwB} was considered for films of thickness $L$ with $(+,+)$ and $(+,-)$ {\bc}s. This is relevant for describing the fluctuation-induced force arising close to the consolute point of a binary liquid mixture. One of the two corresponding species might be preferentially adsorbed at the surfaces of the film. 
An analytic expression in terms of Jacobi elliptic functions was derived for the mean-field order parameter profile $\bar\phi(x_\perp)$ as a function of the coordinate $x_\perp$ across the film and of the reduced temperature $r$. These expressions were used to calculate the value of $T_{ij}$ at the confining walls of the film and thus the {\ccf},  which acts on each of them. 
The important qualitative features of the resulting scaling function $X_{\rm Cas}$ --- confirmed by quantitative, accurate estimates based on {\mc} simulations and by exact results in spatial dimension $d=2$
(see, e.g., Ref.~\citenum{DD2022} 
--- are the following: 
(a) for $(\pm,\pm)$ {\bc}s the force is \emph{attractive} for all (positive and negative) values of the scaling variable
$x$ (see the end of Sec.~\ref{subsec:CE})  and it attains its maximum strength below $T_c$; 
(b) for $(\pm,\mp)$ {\bc}s the force is \emph{repulsive} for all values of the scaling variable $x$ and it attains its maximum strength above $T_c$; 
(c) for a fixed value of the scaling variable
$x$, the attraction for  $(\pm,\pm)$ {\bc}s is generically weaker than the repulsion for $(\pm,\mp)$ {\bc}s. 
In all cases, the force vanishes exponentially for $L \gg \xi$. Here we have assumed that the fluctuating medium in the bulk is in its disordered phase for temperatures $T>T_c$, although also the opposite case can apply to binary liquid mixtures with a lower critical point.  As discussed in Sec.~\ref{ssec:FtoC-patter}, in the case of chemically patterned substrates or of weak surface fields $h_{1,2}$ in Eq.~\eqref{eq:LGWwB},  within this mean-field approximation the {\ccf} loses some of these qualitative features. 

The effective Hamiltonian in Eq.~\eqref{eq:LGWwB}, generalised to a two-component {\op} $\phi(x)$, belongs to the same universality class as the XY lattice model, and therefore it describes the transition from normal to superfluid $^4$He. In this case, as anticipated above, the natural {\bc}s at the surfaces of the film are Dirichlet ones, because of the vanishing of the superfluid {\op}  outside the film. In the actual realisation of the  film,\cite{ChanChap-23} it is spatially confined between a metallic substrate and its vapor, within which the {\op}  vanishes. 
The mean-field analyses of this effective Hamiltonian --- which was done both within the field-theoretical approach discussed here\cite{ZSRKC2007,MGD2007} and by considering suitable lattice models\cite{MGD2007}  --- revealed that the resulting mean-field {\ccf}  takes the scaling form in Eq.~\eqref{eq:FC} with the appropriate mean-field critical exponents. The scaling function exhibits the following qualitative features: 
(a) it vanishes above the bulk superfluid transition temperature $T_\lambda$, i.e., for $x>0$ (see the end of Subsec.~\ref{subsec:CE} with $T_c\mapsto T_\lambda$),
(b) it becomes attractive for $x<0$, i.e., $X_{\Cas} < 0$, 
(c) upon decreasing $x<0$ (i.e., reducing $\xi$)  $X_{\Cas}$ develops a sharp and pronounced minimum, (d) followed by an exponential decay as a function of $|x|^\nu \sim L/\xi$.
Some of these qualitative features are confirmed by {\mc} simulations (see Sec.~\ref{ssec:SF-num}) and are borne out experimentally (see, e.g., Ref.~\citenum{ChanChap-23}). 
The presence of a pronounced minimum belongs to those features of  $X_\Cas$ which are valid beyond mean-field theory, while the fluctuations cause $X_\Cas$ to become nonzero for $T>T_\lambda$, and to decay exponentially upon increasing $T \gg T_\lambda$, whereas it attains a nonzero value\cite{ZSRKC2007} for $T\ll T_\lambda$ with, inter alia, a contribution due to capillary-wavelike surface fluctuations\cite{ZRK2004} if the film is bounded on one side by a fluid-vapor interface.

While typically the mean-field predictions mentioned above are modified by statistical fluctuations below the upper critical dimensionality $d_c=4$ of the transition under consideration, there are experimentally relevant instances in which this is not the case and the mean-field predictions turn out to be (up to logarithmic corrections) quantitatively reliable. 
This holds, e.g., for the tricritical end point\cite{KD92b}, which emerges in the phase diagram of $^3$He-$^4$He mixtures when the line of the continuous normal-to-superfluid transitions joins the critical point of demixing at the top of the two-phase coexistence region between a $^4$He-rich superfluid phase and a $^3$He-rich normal phase (see, c.f., Fig.~\ref{fig:mix}$(a)$).
The {\ccf}s arising at this tricritical point, were investigated experimentally\cite{B99,GC2000,GC2000b,GC2002,UBMCFR2003,IB2005,UBMCR2003,UB2004}, leading also to a quantitative determination\cite{GC2002} of $X_\Cas$. (As mentioned further below, $X_\Cas$ near the tricritical point actually depends on two scaling variables.)
Interestingly enough, the force turned out to be \emph{repulsive}, a fact which was theoretically rationalised via a suitable microscopic model\cite{MD2006}  (the vectoralised Blume-Emery-Griffiths model)  in terms of the emergence of  $(O,+)$ 
{\bc}s  for that specific experimental setting. In particular, the $O$ {\bc} is realised at the solid substrate on which the wetting films are formed while the $+$ {\bc} turns out to be effectively realised at the liquid/vapour interface.\cite{MD2006} 
In order to describe the tricritical behaviour, the bulk contribution in Eq.~\eqref{eq:LGWwB} has to be modified as to include the term $\vv\, (\phi^2)^3/6!$, 
where the mean-field tricritical point is accessed by a suitable tuning such that $u=0$. This is physically achieved by tuning the $^4$He-$^3$He concentration to its critical value. 
The possible deviation of $u$ from zero actually introduces a further scaling variable $u L$ which $X_\Cas$  depends on in addition to $x$ introduced in the last but one paragraph of Sec.~\ref{subsec:CE}. The latter is expressed here as $r L^2$ in terms of the parameters in Eq.~\eqref{eq:LGWwB},  as appropriate for the mean-field nature of the tricritical point\cite{MGD2007} (up to logarithmic corrections).
The tricritical Casimir force can then be calculated as explained above, i.e., within the mean-field approximation either of Eq.~\eqref{eq:LGWwB} modified as discussed or of the lattice model mentioned above.\cite{MD2006,MGD2007} This yields predictions for  $X_\Cas$ in good agreement with the available experimental data. This is illustrated in, c.f., Fig.~\ref{fig:mix} discussed in Sec.~\ref{sec:exp}.
The logarithmic corrections in $d=3$ actually imply a residual dependence on the film thickness $L$ of the algebraically rescaled tricritical Casimir force at the tricritical point.\cite{MD2006,MGD2007} These corrections are due to the fact that the physical spatial dimensionality $d=3$ coincides with the upper critical dimensionality $d_c$ of a tricritical point. This coincidence does not occur for common critical points, for which $d_c=4$ and for which logarithmic corrections are experimentally out of reach. In this respect, tricritical Casimir forces may provide a way to probe  these corrections directly. Note, however, that the limited experimental data discussed in, c.f., Sec.~\ref{subsec:EIE} (see also Fig.~\ref{fig:mix}) do not allow for such a detailed analysis.

In addition to the Landau-Ginzburg-Wilson functional in Eq.~\eqref{eq:LGWwB}, one can also consider effective functionals to be minimised (as in the case of the mean-field approach described above) which are constructed in such a way that the values of the bulk critical exponents are close to those determined numerically in the physically relevant spatial dimension $d=3$. 
This approach, introduced  in 
Refs.\!\nocite{FU90a}\citenum{FU90a} and \!\nocite{FU90b}\citenum{FU90b}, 
can also be successfully used in order to estimate   $X_{\rm ex}$ for the Ising universality class in a slab in $d=3$ with symmetry-breaking {\bc}s.\cite{BU98,BU2008,UB2013}

\subsubsection{Effects of fluctuations}

The approaches discussed above --- i.e., determining either the finite-size free energy or the expectation value of the stress tensor at a suitable surface --- can be readily generalised to the case in which statistical fluctuations around $\bar\phi(\x)$ are considered.  In both cases, accounting for fluctuations requires a suitable field-theoretical renormalization of the original effective Hamiltonian $\HH[\phi]$ which, however, proceeds as in the bulk, i.e., in the absence of confinement.\cite{KD92a}

Within this field-theoretical framework (as well as in the cases of exact solutions of lattice models and the corresponding {\mc} simulations) it is actually rather natural to consider this way also 
{\bc}s, such as, inter alia, periodic and antiperiodic ones, or to analyze surface
multicriticality. However, these specific characteristics are of limited interest for the description of directly accessible experimental systems. 
Yet, these kinds of studies are highly valuable in order to rationalise the results of certain pertinent Monte Carlo
simulations\cite{KL96} of lattice models, and for highlighting possible limitations\cite{DGS2006,GD2008} of some of the analytical approaches mentioned above. These systems allow one also to study crossover phenomena\cite{SD2008,H2011} and the role of
surface universality classes in determining the attractive versus the repulsive character
of the associated {\ccf}.
These field-theoretical investigations extend in
several respects certain exact results available in 
$d=1$\cite{DD2022,RZSA2010} and $d=2$\cite{DD2022,C86,BCN86,A86,ES94,RZSA2010,AM2010}
for various bulk universality classes. These results are based either on the exact solution of suitable lattice models or on the consequences of conformal invariance at the critical point.

Rather generically, if the bulk symmetry of the unconfined model is broken explicitly by the 
{\bc}s (as, e.g., in the case of surfaces with $(\pm)$ {\bc}s and zero bulk field) or 
if the temperature is within the range of values for which such a symmetry is spontaneously
broken, a non-trivial mean-field {\op}  profile $\bar\phi(\x)$ emerges. 
In this case, analytical progress beyond mean-field theory\cite{K97,ZSRKC2007,MGD2007} is significantly impeded and requires a considerable effort\cite{K97}.
Accordingly, except for those cases in which the mean-field or Gaussian
approximations are quantitatively reliable, such as at the tricritical point of
$^4$He-$^3$He mixtures\cite{MGD2007} discussed above or for understanding
the low-temperature fluctuations\cite{ZRK2004} of $^4$He, 
a quantitatively reliable analysis of the experimentally relevant case of confined systems in $d=3$ has to rely on {\mc} simulations of suitable, so-called improved, lattice models (see Sec.~\ref{ssec:SF-num}).

As an illustration of a field-theoretical calculation of the scaling function $X_\Cas$ of the {\ccf} in an experimentally relevant context, we consider the case of the superfluid transition, described by Eq.~\eqref{eq:LGWwB} with a two-component {\op} and with $(O,O)$ {\bc}s, As mentioned in Sec.~\ref{sssec-MFT}, the mean-field scaling function $X_\Cas$ vanishes for $T>T_\lambda$. Therefore the contribution to the force is determined only by statistical fluctuations beyond the mean-field approximation. The corresponding prediction can be read off by specializing the expressions presented in 
Ref.~\citenum{KD92a}.
It follows from the calculation of the renormalised one-loop free energy for a variety of pairs of {\bc}s with $\bar\phi=0$ (i.e., periodic, antiperiodic, Dirichlet, Neumann). This accounts for Gaussian fluctuations. In particular, it turns out that (see Eq.~(6.6) in Ref.~\citenum{KD92a}) 
\begin{equation}
\label{eq:Xcas-XYg}
X_\Cas^{(O,O)}(y_+) = - \frac{2}{(4\pi)^{(d-1)/2}} \frac{y_+^d}{\Gamma((d+1)/2)}\int_1^\infty \!\!\rmd x\, \frac{(x^2-1)^{(d-1)/2}}{\rme^{2 x y_+}-1},
\end{equation}
in terms of the scaling variable $y_+ = L/\xi_+$, where $\xi_+$ is the so-called second-moment correlation length above the critical temperature, while $d$ is the spatial dimension of the confined system. This one-loop result is exact for $d=4$, because the only relevant fluctuations are the Gaussian ones considered here. The value of the scaling function at the critical point, i.e., for $y_+=0$, is the so-called \emph{critical Casimir amplitude} $\Delta^{(O,O)}$, which was calculated in Ref.~\citenum{KD92a} up to two-loop order (see Eq.~(5.16) therein) where the lowest-order perturbative correction $\propto u$ was consistently evaluated at the Wilson-Fisher fixed-point $u=u^*$ for the present case of a two-component {\op}  with $u^* = 3\varepsilon/10 + O(\varepsilon^2)$; $\varepsilon = 4-d$ is the deviation from the upper critical dimensionality $d_u = 4$. The scaling function in Eq.~\eqref{eq:Xcas-XYg} corresponds to an attractive force, the modulus of which decreases from the value $|\Delta^{(O,O)}|$ at $y_+=0$ and which vanishes  exponentially upon increasing $y_+$.
Interestingly, universality implies that the scaling function in Eq.~\eqref{eq:Xcas-XYg}, calculated within the Gaussian approximation of Eq.~\eqref{eq:LGWwB}, is actually the same\cite{GD2006} as the one of the {\ccf}  which arises in a confined ideal Bose gas in the grand canonical ensemble upon approaching the Bose-Einstein condensation point for vanishing chemical potential $\mu=0$; similarly, the same applies in the presence of interactions. This can be shown explicitly by a direct analytic calculation\cite{MZ2006} of the grand canonical potential of the confined gas, based on elementary statistical mechanics\cite{H87} (i.e., without resorting to the field-theoretical model in Eq.~\eqref{eq:LGWwB}) and by taking into account the quantization of the free particle wavevectors due to the Dirichlet {\bc}s at the walls. From this grand potential, one can infer the finite-size contribution according to Eq.~\eqref{eq:Fdec} and thus the scaling function of the {\ccf} in terms of the correlation length $\xi_+(\mu\to 0^-)  \propto (-\mu)^{-1/2}$.
Finally, we mention that significant analytical insight into the emergence of {\ccf}s in confined statistical systems has been provided 
by the solution of the exactly solvable models reviewed in Ref.~\citenum{DD2022}, e.g., 
the spherical model
\cite{D93,D96,D98,CD2004,DG2009} with various {\bc}s. 

\subsection{Numerical approaches} 
\label{ssec:SF-num}

As anticipated above, exact predictions for the scaling function of the {\ccf}  are available in $d=2$ (based on the solutions of either lattice models or field theories due to conformal invariance, as well as for various {\bc}s and geometries of the boundaries) or in general dimensions $d$ for simplified models which actually do not belong to the universality classes of experimentally accessible systems. 
Instead, predictions for the corresponding universality classes in the physically relevant case $d=3$, can be obtained via a perturbative expansion (as discussed in Sec.~\ref{ssec:SF-ana}), a derivative expansion,\cite{JNa2013,ENS2016} or within effective models\cite{BU98,BU2008,UB2013}, which necessarily involve approximations of various nature. This implies that, in general, \emph{quantitative} predictions for $X_\Cas$ are beyond the reach of these approaches, while they can be obtained efficiently by {\mc} simulations of lattice models which belong to the universality classes of interest or, to a lesser extent, by molecular dynamics simulations. 

In particular, $X_\Cas$ for the demixing transition of a classical binary liquid mixture can be determined  by simulating  a three-dimensional slab of the lattice Ising model (or improvements thereof, see below) with fixed spins at the boundaries, as to realise the
physically motivated symmetric $(\pm,\pm)$ or antisymmetric $(\mp,\pm)$ 
{\bc}s.\cite{VGMD2007,VGMD2009,H10b,H2014} 
Similarly, for the superfluid transition in $^4$He, $X_\Cas$ can be conveniently determined by simulating  a
slab of the XY model (or improvements thereof, see below) with free {\bc}s, which realise $(O,O)$ 
{\bc}s.\cite{VGMD2007,VGMD2009,H2007,H2009,H2010,H2009b}

However, extracting the {\ccf}  from {\mc} simulations requires the development of suitable approaches. In fact, 
simulations of lattice models in thermal equilibrium are particularly efficient for determining the expectations values of \emph{local} quantities, which can be expressed in terms of a small number of local degrees of freedom on the lattice. Unfortunately, the {\ccf}  acting on a surface cannot be expressed directly  in such a form, given that the natural observable for achieving this, i.e., the stress tensor discussed in Sec.~\ref{sssec-MFT} for a field theory, cannot be generalised to lattice systems (and discrete degrees of freedom) apart from very special cases\cite{DK2004}.

Accordingly, in order to determine the {\ccf}, one has first to calculate numerically the free energy  $\Omega(\beta,L,A)$ of a film of transverse area $A$, thickness $L$, and in equilibrium at temperature $\beta^{-1}$. 
(Compared to the notation  introduced in Sec.~\ref{subsec:CE} for the finite-size free energy $\Omega_V$, we omit hereafter the indication of the subscript $V$ in order the simplify the notation.)
Then, from the dependence of 
$\Omega(\beta,L,A)$ on $L$ one infers the force according to Eqs.~\eqref{eq:Fdec} and \eqref{eq:FC}.  
With this approach one still faces the problem that $\Omega(\beta,L,A)$ cannot be directly expressed as the statistical average of a suitable local quantity, which lends itself to numerical simulation. 
However, its variation $\Delta \Omega$ as a function of $L$ (e.g., $\Delta \Omega (\beta,L,A) = \Omega (\beta,L,A) - \Omega (\beta,L-1,A)$ or analogous, discrete estimates of the derivative with respect to $L$, measured in units of the lattice spacing), can be actually expressed\cite{H2007,VGMD2007,H2009,ENS2014} as an integral over an auxiliary, dimensionless parameter $\lambda$ which the statistical average of a suitable local quantity depends upon.
In particular, depending on the physical significance of $\lambda$ and of that local quantity, three general strategies have been considered:
%
%

(i) $\lambda \in [0,1]$ controls the continuous interpolation between the Hamiltonians $H_1$ and $H_2$ of the system consisting, respectively, of the model in the film geometry with thickness $L$ and of the model in the film with thickness $L-1$ plus an isolated layer of ``thickness'' 1 (see, e.g., Fig.~1 of Ref.~\citenum{VGMD2007}), such that both $H_1$ and $H_2$ involve the same number of degrees of freedom. The free energy difference $\Delta \Omega$ can then be expressed as  $\int_0^1\rmd \lambda \langle H_2 - H_1\rangle_{H_{\rm cr}(\lambda)}$ where the expectation value $ \langle \cdots \rangle_{H_{\rm cr}(\lambda)}$ is the canonical one at the temperature $\beta^{-1}$ with the interpolating crossover Hamiltonian $H_{\rm cr}(\lambda) = (1-\lambda)H_1 + \lambda H_2$. This strategy was introduced in Ref.~\citenum{VGMD2007} and used in order to investigate various bulk and surface universality classes\cite{VGMD2009,RSVD19}, 
also in the presence of variable\cite{VMD2011} or random\cite{MVDD2015,PT2015} surface fields. 
It was also extended in order to calculate the {\ccf} in the presence of bulk ordering fields.\cite{VD2014}  
This approach was also used for investigating the {\ccf}  emerging with patterned substrates\cite{TD2010,TD2010rev,ParisenToldin2013,PTD2015} as well as with a more complex shape of one of the confining surfaces\cite{VM2015,VED2013}, as discussed further below.
Alternatively, a similar approach consists of considering an interpolation between $H_1$ and $H_2$ which is more sophisticated compared to $H_{{\rm cr}}(\lambda)$. For example one starts from $H_1$ corresponding to a film of thickness $L-1$ and adds one by one the degrees of freedom which are necessary in order to add a complete layer in order to form the film of thickness $L-1$ (see Ref.~\citenum{H2009}). In this case, $\Delta \Omega$ can be expressed as 
$\sum_{i=1}^N \ln \langle \exp[-\beta(H^{(i+1)}-H^{(i)})]\rangle_{H^{(i+1)}}$ where $\{H^{(1)}, H^{(2)}, \ldots H^{(N)}\}$ is the sequence of Hamiltonians involved in the interpolation, and the canonical expectation value $\langle \cdots \rangle_{H^{(i+1)}}$ is taken with the Hamiltonian $H^{(i+1)}$ (see Ref.~\citenum{H2009} for further details). 
%
%

(ii) In the second strategy, $\lambda$ is identified with the inverse temperature $\beta$, and the thermodynamic 
relationship\cite{H87} $\U = \partial (\beta \Omega)/\partial \beta$
between the (finite-size) free energy $\Omega$ and the (finite-size) internal energy $\U$ is used in order to calculate $\Delta\Omega$. (Here it is assumed that the remaining thermodynamic parameters of the system are kept constant.) 
In fact, based on the numerical  determination of $\Delta\U\equiv \U(\beta,L,A)-\U(\beta,L-1,A)$ for a suitable set of values of temperatures, it is possible to provide an accurate numerical estimate of the integral of $\Delta \U$ over $\beta$, which is required in order to invert the previous relationship between $\U$ and $\Omega$ and therefore between $\Delta\U$ and $\Delta\Omega$.
In turn, $\Delta \U$ and $\U$, from which $\Delta\Omega$ is derived, can be directly expressed as the canonical expectation value of the Hamiltonian $H_1$ of the lattice system, which is amenable to {\mc} simulations.
This approach was proposed in Ref.~\citenum{H2007} and applied for studying various bulk and surface universality classes\cite{H2009b,H2010,H10b,H2012} as well as crossover phenomena in the latter.\cite{H2011} 
Similarly, this approach was also used beyond the film geometry\cite{H2013,H2015}.
%
%

(iii) In a magnetic model (e.g., the Ising model as representative of the corresponding universality class), $\lambda$ is identified with the external magnetic field $h$. The free energy $\Omega(\beta,h,L,A)$ of a film of thickness $L$, transverse surface area $A$, at inverse temperature $\beta$, and in the presence of such a non-vanishing field $h$, is obtained from that $\Omega(\beta,h_0,L,A)$ of a reference state with field $h_0$ via thermodynamic integration: $\Omega(\beta,h,L,A) = \Omega(\beta,h_0,L,A) + \int_{h_0}^h\rmd h' M(\beta,h',L,A)$, where $M(\beta,h',L,A)$ is the magnetization of the system which is amenable to {\mc} determination. The value of $|h_0|$ can be chosen to be sufficiently large so that the corresponding correlation length is small, and whereby $\Omega(\beta,h_0,L,A)$ coincides with the free energy which a volume $L\times A$ of the system would have in the \emph{bulk}.  Thus the critical Casimir contribution is encoded in the integral, together with a surface contribution. Eventually, the scaling function of the free energy can be extracted from the values of a suitable estimate of these integrals.
This approach was first used in Refs.~\citenum{ENS2014} and \citenum{V2014} for obtaining the scaling function of the {\ccf}  also in the presence of a bulk field, both in the film geometry\cite{ENS2014} or for two spheres\cite{V2014} (see below).
%
%

An alternative approach for determining the {\ccf}, which does not require a computation of $\Delta\Omega$ and which is, in spirit, close to what is pursued in certain experimental settings (see Sec.~\ref{sec:exp}), was proposed and used in Ref.~\citenum{HH2014}. In particular, one considers a wall in $d=2$ (i.e., a line of the lattice where the degrees of freedom are fixed according to the {\bc}s) which is immersed in a critical medium confined to a strip, with boundaries parallel to the wall and certain {\bc}s. The wall separates the systems into two adjacent sub-strips to which the medium is confined. Therefore the wall is subject to a bulk pressure and a {\ccf}  from both sides. The rigid wall is assumed to be mobile, such that its position $h$ across the strip is allowed to fluctuate. Accordingly, the {\mc}  simulation of the whole system includes also moves in which the wall interchanges its position with one of the two neighboring lines of degrees of freedom (such as the spins for the Ising model considered in Ref.~\citenum{HH2014}), with suitable flipping rates which satisfy detailed balance. As a consequence, the statistical distribution $P(h)$ of $h$ --- which can be measured numerically --- is determined by the potential $\Phi(h)$ of the total force acting on the wall via the relation $P(h) \propto \exp\{-\beta \Phi(h)\}$; thus $\Phi(h)$ follows from $P(h)$. The latter contains the contribution of the {\ccf},  which the medium confined to each sub-strip exerts on the wall, and, in principle, it allows one to infer the scaling function of the force\cite{HH2014}. 
This approach was also extended to the case involving a spherical particle.\cite{HH2014}

The numerical determination of   $X_\Cas$ is also affected by finite-size corrections to the leading finite-size scaling behaviour of various origin. Accordingly, it is important to devise strategies which reduce them. In this respect, it is particularly convenient to focus on lattice models, which belong to the universality class of interest, but which have been modified by the introduction of irrelevant terms in their Hamiltonian, which allow one to choose them as to minimise corrections to scaling and thus corrections to finite-size scaling. 
These so-called improved models --- widely used for an accurate determination of bulk critical exponents\cite{PV2002}  --- have been also used to study the {\ccf}  for the Ising 
universality class, represented by a suitable Blume-Capel model\cite{H10b,H2011,H2012,TD2010,TD2010rev,ParisenToldin2013,PTD2015,PT2015,H2015},
and  the XY universality class, represented by Eq.~\eqref{eq:LGWwB} for a two-component {\op}  and a suitable value\cite{H2009b,H2010}  $u^*$ of the coupling constant $u$.

Concerning the Ising universality class, an alternative to the various approaches mentioned above consists of performing a molecular dynamic simulation of a model binary fluid confined within a slab and of determining the pressure which the fluid exerts on the confining walls in the semi-grand-canonical ensemble\cite{Puosi2016}. Upon approaching the second-order demixing transition one observes the emergence of a contribution to the pressure on the walls, which is due to the {\ccf}  and which, in principle, can be used to extract the corresponding scaling function. However, in comparison with {\mc} simulations, this method still suffers from sizeable finite-size corrections, because the system sizes which can be actually simulated are rather limited by the high computational effort. Similarly, the number of possible {\bc}s which can be realised is reduced. However, this approach, if developed further, could provide access to the non-equilibrium properties of the {\ccf}  (see Sec.~\ref{sec:Dyn}).
Alternatively, instead of considering a binary fluid, one might focus on a medium composed of identical, interacting depletant particles in the presence of two spatially fixed colloidal particles (see, c.f., Sec.~\ref{sec:FtoC}), which are amenable to Monte Carlo simulations.\cite{GZTS2012} The bulk phase diagram of the medium as a function of temperature $T$ and volume fraction $\phi$ of the depletant particles features phase separation and a critical point. The force acting on the fixed colloidal particles can be calculated by performing virtual displacements, as discussed in Ref.~\citenum{GZTS2012}. From this force one can derive the effective interaction potential between the colloids and thus, upon approaching criticality, the potential of the {\ccf}s.

The {\ccf}  can be studied, inter alia, also numerically by exploiting the well-known relationship between the equilibrium behaviour of quantum and classical statistical models.\cite{ENS2016}  
This connection implies that a two-dimensional \emph{quantum} system with surface area $A$ in thermal equilibrium and at the inverse temperature $\beta$ has the same partition function (and thus the same statistical properties) as the corresponding \emph{classical} model in a three-dimensional system of size $A\times\beta$ and periodic {\bc}s along the additional dimension of linear extension $\beta$. The temperature $T$ of the corresponding classical system is determined solely by the parameters entering the Hamiltonian of the two-dimensional quantum system and is not related to the temperature $\beta^{-1}$ of the original quantum system.
Accordingly, reversing the argument, the free energy $\Omega$ 
(derived from the partition function) of a three-dimensional classical film of thickness $L$ (and transverse area $A$) depends on $L$ in the same way as the free energy of the corresponding two-dimensional quantum system (of area $A$) does on the inverse temperature $\beta=L$. (For simplicity, here we use the units such that $h = c =1$.)
The {\ccf}  can then be extracted from the thermodynamic relationship $\partial \Omega/\partial L = T \, \U_q(\beta=L)$, where $\U_q(\beta)$ is the internal energy of the two-dimensional \emph{q}uantum system in equilibrium and at the inverse temperature $\beta$.
In turn, the latter can be determined, via a variety of approaches developed for studying quantum systems and quantum phase transitions, encompassing analytical and numerical approaches. From the perspective of numerical studies, which we are concerned with in the present section, we mention the possibility to use quantum {\mc} methods in $d=2$, which turn out to confirm the results of classical {\mc} simulations for Ising films in $d=3$.\cite{ENS2016} 
However, beyond its conceptual appeal, this approach seems to be limited to the case of periodic {\bc}s at the surfaces of the film, which unfortunately lacks experimental relevance.\\[-3mm]

\emph{Beyond the homogeneous film geometry.---} In the above overview of the numerical investigation of {\ccf}s we have focussed primarily on the film geometry with homogeneous preferential adsorption along its long axis.
However, as we shall discuss in Sec.~\ref{sec:FtoC}, experimental settings often involve curved surfaces (such as those of colloidal particles) and chemically patterned substrates. These cases have also been studied numerically by using the various approaches mentioned above, possibly accompanied by the development of tailored algorithms. 

In particular, the case in which one of the two confining surfaces of the film is chemically patterned with a single chemical step\cite{PTD2015} or with a periodic pattern of alternating adsorption preferences\cite{TD2010,TD2010rev,ParisenToldin2013} (realizing alternatively $(+)$ and $(-)$ {\bc}s) was considered, with the other surfaces imposing $(+)$ or $O$ {\bc}s. 
For an isolated step the line contribution to the {\ccf} was determined as the necessary information needed to investigate a pattern with well-separated steps. 
For a generic pattern, in addition to the scaling variable $y \equiv L/\xi$ in Eq.~\eqref{eq:FC},  or equivalently to $x$ (see end of Subsec.~\ref{subsec:CE}), the scaling function $X_\Cas$ of the {\ccf} depends additionally on scaling variables related to the patterning, i.e., $S_\pm/L$ where $S_+$ and $S_-$ are the widths of the stripes with $(+)$ and $(-)$ {\bc}, respectively (with $S_+=S_-$ in Refs.~\citenum{TD2010}, \citenum{TD2010rev}, and \citenum{ParisenToldin2013}). Depending on the {\bc}s at the homogeneous surfaces,  a change of the force from attractive to repulsive may occur upon varying $x$ with the other scaling variables kept fixed.
If, instead, the medium is confined by a surface with a certain preferential adsorption, say $(+)$, and another one with a quenched, random local adsorption preference which can be either $(+)$ or $(-)$, one observes a crossover between an  attractive case $(+,+)$ and a repulsive one $(+,-)$, with a crossover function which depends on the disorder strength via a suitable scaling variable.\cite{PT2015,MVDD2017}

The numerical investigation of {\ccf}s beyond the film geometry can be carried out on the basis of the approaches mentioned above, although one might need to generalise standard algorithms, such as the cluster algorithm\cite{H2013,H2015,HH15}, or to consider lattices with adaptive spacing\cite{H2015b}.  
In particular, the various geometrical arrangements, which were investigated numerically, include in $d=2$, circular particles\cite{HH2014} or elongated ones (``needle'')
within a strip\cite{VED2013}, cubic particles within a film in $d=3$\cite{VM2015}, spheres close to a plane\cite{H2013}, two spheres\cite{V2014} in the bulk of the system, and 
more circles in $d=2$\cite{HH15}. All aforementioned studies focused on the Ising universality class with $(\pm)$ {\bc}s at the involved surfaces. 


\section{From films to colloids}
\label{sec:FtoC}

As discussed in the previous section, the emergence of {\ccf}s was originally studied both theoretically and experimentally (see, c.f.,  Sec.~\ref{sec:exp}) in the film geometry. However, in view of possible applications to soft matter as well as for a direct measurement of these forces, it turned out to be useful to consider geometries involving spherical particles, such as colloids: in fact, the thermal energy scale, which controls the strength of {\ccf}s, is the same as the one of the interactions which are typically at play in soft matter; accordingly, they are expected to have a measurable impact on the observed behaviour of, e.g., colloidal suspensions. 
Rather generally, upon reducing further the degree of symmetry of the considered geometry, additional components of the {\ccf}   arise compared to the case of homogeneous films or of a homogeneous spherical particle close to a homogeneous flat surface.

The change of the geometry of the confining surfaces poses significant challenges for the theoretical study of the critical Casimir effect via the analytical approaches discussed in Sec.~\ref{ssec:SF-ana}.  Hardly any analytic progress can be made as soon as the {\bc}s are imposed on non-planar surfaces, and the geometry is more complex than that of films. Within mean-field theory (see Sec.~\ref{sssec-MFT}), some advances are still possible concerning the numerical minimization of the functional in Eq.~\eqref{eq:LGWwB} with modified boundaries, and the determination of the {\ccf}  via the stress tensor. 
This was carried out, for example, in the case of a spherical\cite{HSED98} or ellipsoidal\cite{KHD2009} particle approaching a flat and homogeneous surface, as well as in the cases of films in which one or both confining surfaces have a sawtooth cross section\cite{THD2008} or one is a crenellated surface\cite{THD2015}. Similarly, within the mean-field approximation it was possible to study the case in which a cylindrical colloidal particle approaches the soft and therefore responsive interface separating the two liquid phases in one of which the colloid is immersed.\cite{LHTD2014}
As an exception, analytical and reliable predictions can be formulated also beyond the film geometry for systems which are exactly at their critical point (i.e., formally $\xi=\infty$). 
For such systems, conformal invariance (see, e.g., Ref.\!\nocite{BCN86}\citenum{BCN86}) allows one to map geometries with non-planar boundaries (specifically spheres) into the film one. 
This enables one to calculate analytically the forces acting on the former in terms of those acting on films.\cite{BE95} 
In $d=2$ this approach can be extended further to boundaries with a modulated adsorption preference\cite{DSE2017}, which are also discussed, c.f., in Sec.~\ref{ssec:FtoC-patter}.
As we mentioned at the end of Sec.~\ref{ssec:SF-num}, the {\ccf} acting on non-planar boundaries can be studied via {\mc} simulations. Nonetheless, the complexity of the problem rises due to the increased number of scaling variables the scaling function eventually depends on.

In fact, in addition to the mesoscopic lengthscales which are relevant in the film geometry, i.e., the correlation length $\xi$ and the surface-to-surface distance $L$ (i.e., the film thickness), the presence of a spherical particle introduces its radius $R$ as the sole parameter describing the geometry of the surface. Accordingly, one generically expects that the {\ccf}  $F_\Cas$, which acts on a sphere and which is due to the effective confinement of the medium when the sphere approaches a plane at a closest distance $L$, depends additionally on the scaling variable $\Delta \equiv L/R$ such that Eq.~\eqref{eq:FC} generalises  to\cite{HSED98}
\begin{equation}
F_\Cas = \frac{k_{\rm B}T}{R} X_{\circ|}(y=L/\xi,\Delta = L/R).
\label{eq:FC-s-p}
\end{equation}
In the case, for example, of two spherical colloidal particles with different radii $R_1$ and $R_2$, the {\ccf}  will have a scaling form as the one above, but with an additional dependence on the dimensionless scaling variable $R_1/R_2$. An analogous extension is expected, e.g., when the particle has a spheroidal shape and it approaches 
a planar surface or another (spheroidal) particle. 

The scaling function $X_{\circ|}$ of two variables, which appears in Eq.~\eqref{eq:FC-s-p}, as well as the dependence of the scaling function $X_\Cas$ in Eq.~\eqref{eq:FC} on the additional scaling variables related to possible patterns of the film surfaces (see the end of Sec.~\ref{ssec:SF-num}), can in principle both be determined by numerical simulations of suitable lattice models, as anticipated in Sec.~\ref{ssec:SF-num}. 
However, this might be computationally demanding due to the increased number of variables involved. Accordingly, it is often useful  to consider first some relevant limits. In particular, the case of small particles, formally corresponding to the limit $\Delta \to \infty$, can be addressed, on the basis of the knowledge of the critical behaviour in the presence of boundaries\cite{D86,D1997}, via the so-called \emph{small-sphere expansion}\cite{HSED98}.
This approach can also be extended to non-spherical small particles such as dumb-bells and needles\cite{E2004,VED2013}, and also to chemically inhomogeneous (Janus) particles in $d=2$\cite{SEMD20}.
The opposite limit $\Delta \to 0$, instead, can be addressed on the basis of the available information concerning the {\ccf}  acting between parallel plates. This is accomplished by using the so-called \emph{Derjaguin approximation}, 
also known as \emph{proximity force approximation}\cite{D34,I2011,P2006}, which assumes pairwise additivity of the fluctuation-induced forces. Since most of the available experimental data fulfil the limit $\Delta \ll 1$, here we focus primarily on this case.

\subsection{Pairwise additivity} 
\label{ssec:FtoC-DJ}

Within the Derjaguin approximation, two non-planar surfaces, which approach each other, are completely decomposed into a set of pairs of opposing and parallel surface elements of infinitesimal area $\rmd A$; the two elements of such a pair belonging to the two different surfaces are separated by a distance $L(\rmd A)$. Each of these pairs is assumed to provide a contribution $\rmd F$ to the force. The latter is the {\ccf}  acting on a surface of extension $\rmd A$, which belongs to the confining surfaces of a film of thickness $L(\rmd A)$, given by Eq.~\eqref{eq:FC}. 
The total force $F_\Cas$ is calculated as the integral over all these contributions under the assumption of pairwise additivity.
Accordingly, the knowledge of the scaling function $X_\Cas$  for two macroscopic parallel plates allows one to predict the force acting in the presence of non-planar boundaries, such as those formed by a spherical particle approaching a flat surface, another spherical particle, or a topographically patterned but chemically homogeneous substrate. As it will be discussed in more detail in Sec.~\ref{ssec:FtoC-patter}, the Derjaguin approximation allows one to study also the cases of surfaces with spatially varying preferential adsorption as well as 
of non-spherical particles. 
In all these instances one can take advantage of the quantitatively accurate determination of the {\ccf}  inside homogeneous films with the appropriate {\bc}s, obtained, e.g., via {\mc} simulations.

For the concrete and experimentally relevant case of a spherical particle approaching a planar surface, the Derjaguin approximation is expected to be valid when the radius $R$ of the sphere is significantly larger than the distance $L$ of closest approach between the particle and the surface.  In Eq.~\eqref{eq:FC-s-p} this  corresponds to the limit $\Delta = L/R\to 0$.
Within this approximation, it turns out that $X_{\circ|}(y,\Delta\to 0) \rightarrow \Delta^{-2}X^{({\rm DA})}_{\circ|}(y)$, 
where $X^{({\rm DA})}_{\circ|}(y)$ can be expressed in terms of a certain elementary integral of the scaling function $X_\Cas(y)$ for the {\ccf}  in the film geometry [Eq.~\eqref{eq:FC}].  The potential $\Phi_\Cas(L)$ of the resulting {\ccf}   --- which is typically the quantity accessible to experiments (see Sec.~\ref{sec:exp}) ---  turns out to be given, within the same approximation, by
\begin{equation}
\frac{\Phi_\Cas(L)}{k_{\rm B}T} = \frac{R}{L} \vartheta_{\circ|}(L/\xi).
\label{eq:ccpot-da}
\end{equation}
The scaling function $\vartheta_{\circ|}$  follows from an integration of $X^{({\rm DA})}_{\circ|}(y)$ and inherits the universal features of $X_{\circ|}(y,\Delta)$ (with $\Phi_\Cas(L\to \infty)=0$). Analogously, in the case of spheroidal particles, which are characterised by a set $\{ R_i\}$ of radii,  one expects the Derjaguin approximation to become more accurate for sufficiently small ratios $\{  \Delta_i \equiv L/R_i \}$. 
In particular, in Refs.~\citenum{HHGDB2008} and \!\nocite{GMHNHBD2009}\citenum{GMHNHBD2009} the Derjaguin approximation was used for predicting 
  $ \vartheta_{\circ|}(x)$ on the basis of the {\mc} simulation data for the scaling function in the film geometry (see Sec.~\ref{ssec:SF-num}), which in turn was found to be in agreement with the corresponding experimental data. The qualitative features, which emerge in the case of homogeneous preferential adsorption at the sphere and at the planar surface of the type $(\pm)$, are the same as those for the corresponding {\bc}s in the film geometry. This implies that the force is attractive for equal {\bc}s and repulsive otherwise, while a richer behaviour is expected for colloids exposed to patterned substrates, which are discussed in Sec.~\ref{ssec:FtoC-patter}.

The range of validity of the Derjaguin approximation can be tested in various ways, for example by deriving, within the same approach, the scaling function of the {\ccf} in the film geometry and the one in the sphere-plate geometry, $X_{\circ|}(y,\Delta)$. The former is then used to calculate    $X^{({\rm DA})}_{\circ|}(y)$, which is then compared with $\Delta^2 \times X_{\circ|}(y,\Delta)$ for decreasing values of $\Delta$. This comparison, based on the results of {\mc} simulations for the Ising universality class in $d=3$\cite{H2013}, shows both for $(+,+)$ and $(+,-)$ {\bc}s, that the Derjaguin approximation is actually accurate even up to $\Delta \lesssim 0.5$ for all values of the scaling variable $x=(T/T_c-1)(L/\xi_0^+)^{1/\nu}$ in the case of $(+,+)$ {\bc}s. Instead, for $(+,-)$ {\bc}s,  the accuracy rapidly deteriorates for $x<0$. A similar comparison can be carried out within mean-field theory, i.e., by using the mean-field predictions for the scaling function in the sphere-plane geometry and those for the film geometry. This comparison was studied for an ellipsoidal particle approaching a plane\cite{KHD2009}, where the ensuing \emph{critical Casimir torque} for a non-spherical particle was investigated in detail. In particular, it turned out that sizeable deviations appear between the two approaches upon increasing the aspect ratio of the ellipsoidal particle. 
On the other hand, for the case of a spherical colloid and a substrate with chemical patterns, as discussed in the next section, the suitably extended\cite{TKGHD2009} Derjaguin approximation within the mean-field approximation turned out to provide rather accurate predictions as long as the typical lengthscale of the patterns is larger than the geometric mean of the particle radius and its distance of closest approach to the substrate\cite{TKGHD2010}. 
However, beyond the aforementioned cases, a detailed investigation of the validity of this approximation is still missing. 
The Derjaguin approximation was also used in order to predict the {\ccf}  acting within the film geometry with one crenellated surface\cite{THD2015} on the basis of the mean-field predictions for a film with two flat confining surfaces. 
A suitable modification of the approximation turned out to provide predictions which are in very good agreement with those of a direct numerical calculation within the mean-field approximation.

In passing, we mention that the results of the experimental investigations of the {\ccf}s available  in the literature so far  (see Sec.~\ref{sec:exp}) agree with the  predictions derived on the basis of the Derjaguin approximation,  which is therefore confirmed within the present experimental accuracy and range of  experimental parameters.

\subsection{From homogeneous to patterned surfaces}
\label{ssec:FtoC-patter}
 
The possibility of altering locally the adsorption preference of a material via suitable surface chemical treatments --- which, for example, turn it from being hydrophobic to hydrophilic --- can be used in order to realise surfaces with varying adsorption preferences for the two components of a binary liquid mixture (one of which is typically water). Accordingly, the effective {\bc}s for the corresponding {\op}  $\phi$ can be spatially modulated between the two cases $(+)$ and $(-)$ discussed above, which realises chemically patterned substrates. 
For the film geometry, involving two confining surfaces with a periodically alternating adsorption preference, the emergence of a \emph{lateral} component of the {\ccf},  in addition to the one perpendicular to the surfaces, has been studied within mean-field approximation\cite{SSD2006} or {\mc} simulations (see Sec.~\ref{ssec:SF-num}).  
Such a component emerges also in the experimentally relevant case in which colloidal particles with a certain adsorption preference are exposed to these substrates with spatially modulated adsorption preferences. As a consequence, colloidal particles, while diffusing, tend to localise at those portions of the surface which have the same adsorption preference while they are repelled from those parts of the surface with the opposite preference.  
Similarly, spherical colloids subject to inhomogeneous surface treatments and colloidal clusters obtained by assembling colloids with different adsorption preferences, realise effectively (anisotropic) Janus particles, which are subject to a critical Casimir \emph{torque}, even in the presence of a homogeneous substrate.
By combining these various possibilities one can actually control the spatial distribution and orientation of colloidal particles to a degree which can be reversibly controlled by temperature: upon moving away from the critical point, these critical Casimir interactions become generically negligible and the corresponding spatially modulated force vanishes.

Motivated by actual experiments (see Sec.~\ref{ssec:exp-1c}), the case of a single, homogeneous, and spherical colloid with a certain adsorption preference in front of a substrate, which exhibits a chemical step between the same and the opposite adsorption preference, was considered in Ref.\!\nocite{TKGHD2009}\citenum{TKGHD2009}. The force acting on the colloid, as a function of the position of its centre in this geometrical arrangement, was determined both within mean-field theory and by using the Derjaguin approximation discussed in Sec.~\ref{ssec:FtoC-DJ}. The latter takes into account that the various pairs of infinitesimal flat surfaces $\rmd A$, into which the system is effectively decomposed, realise distinct {\bc}s depending on which part of the substrate they belong to. In particular, the available {\mc} data for the force acting within the film geometry were used in this Derjaguin approximation. Due to the intrinsic additivity of this approximation, the previous step case can be used to predict the force also in the presence of an isolated, say $(+)$, stripe in an otherwise $(-)$ substrate, as well as in the case of periodic lateral patterns. As we also discuss in Sec.~\ref{ssec:exp-1c},  in the former case it turns out that the probability distribution of the position of a single colloid close to the striped substrate, which, inter alia, is due to the action of {\ccf}s, is even able to reveal details of this stripe, such as possible irregular boundaries, which would be otherwise unaccessible.\cite{TKGHD2009,TZGVHBD2011} 

Rather generically, if a colloid is sufficiently close to a patterned substrate, their interaction is dominated by the contribution stemming from the area of the substrate and of the colloid located around the corresponding points of closest approach. Accordingly, it might be repulsive or attractive depending on the local adsorption preferences on the substrate, provided the colloid surface is chemically homogeneous.
On the other hand, if the distance between the particle and the substrate increases, the force follows from an average of the contributions stemming from a larger area around the points of closest approach. Accordingly, its character might change from attractive to repulsive or vice versa. 
The fact that within the film geometry the {\ccf}  is ca.~four times stronger in the case of repulsion than in the case of attraction (see Sec.~\ref{sec:SF}), actually helps to facilitate such a change of character of the force. For a suitably designed substrate one is able to find a point of \emph{stable levitation}. This one is controlled by the {\ccf} only, and therefore it is extremely sensitive to changes of temperature\cite{TKGHD2010}. 
These investigations were briefly reviewed in Ref.~\citenum{GD2011}.

In the presence of non-spherical colloids, either homogeneous or inhomogeneous as far as the preferential adsorption is concerned, the interaction with the patterned substrate may involve the emergence of critical Casimir torques which tend to align the colloidal particle with possible stripes on the substate which they are exposed to. The case of cylindrical homogeneous or Janus-like colloids was investigated in Ref.\!\nocite{LTHD2014}\citenum{LTHD2014} using both the Derjaguin approximation and mean-field theory. This  demonstrates that, in principle, these critical Casimir interactions can be exploited in order to achieve a well-defined and reversible alignment of both chemically homogeneous and Janus cylinders.

The self-assembly and phase behaviour of colloidal suspensions, which consist of patchy or Janus particles, are currently under intensive theoretical and experimental investigation. This calls for a deep understanding of the possible effective interactions at play in these systems. As illustrated above, {\ccf}s stand out in this respect because they  provide both attractive and repulsive, reversible, and temperature-controlled interactions, which can be engineered via surface chemical treatments. In particular, in Ref.\!\nocite{LD2016}\citenum{LD2016}, the interaction between a single Janus particle and a laterally homogeneous substrate or a substrate with a chemical step was studied within the mean-field approximation and a suitably modified Derjaguin approximation. Due to the pairwise additivity of this approximation, it can be used to calculate the effective force between two Janus cylinders and between two Janus spheres, depending on their relative distances and orientations. 
Similarly, the effective pair interaction between two identical spherical colloids --- carrying a circular patch of arbitrary size on their surfaces, characterised by an adsorption preference different from that on the rest of the surface --- can be predicted within the same approximation and can be compared with the available experimental results\cite{BND2020}. In addition, a generic relationship between the scaling function of the inter-particle potential and that of the critical Casimir torque emerges, even beyond the aforementioned specific case\cite{BND2020}.

\subsection{Many particles: beyond pairwise additivity and many-body effects}
\label{ssec:FtoC-mb}

Effective interactions in physics are generically expected to display many-body effects: in the presence of many particles the total force acting on one of them is \emph{not} simply given by the sum of its pairwise interactions with the other particles in the system, contrary to what occurs, e.g., for the gravitational interaction or the electrostatic force in vacuum.
Among the various effective interactions relevant for soft matter\cite{I2011}, which a priori are not pairwise additive, also {\ccf}s are expected to exhibit many-body effects. They are out of reach of any theoretical investigation based on the Derjaguin approximation discussed in Sec.~\ref{ssec:FtoC-DJ}. Thus the theoretical studies of them, as currently available, rely primarily on either mean-field approaches\cite{MHD2012,MHD2014} or  numerical simulations\cite{H2013}. (The experimental investigation of three-body effects is discussed in, c.f., Sec.~\ref{ssec:exp-mb}.) 

In particular, many-body effects can be extracted and quantified by subtracting from the force acting on a particle in a certain configuration, the force which would act on it assuming pairwise additivity, starting from the simplest case of three particles immersed in the medium. In the case of two spheres facing a planar and homogeneous flat substrate  immersed in a near-critical binary liquid mixture, which are studied within the mean-field approximation, three-body effects appear both for the normal and the lateral {\ccf} acting on one of the colloids\cite{MHD2012}. Depending on the relative geometrical arrangement of the two colloids and the substrate, as well as on their {\bc}s and the deviation from the critical temperature of the solvent, a number of interesting features emerge, including a change of sign of the force. As  expected heuristically, the many-body contribution is quantitatively more relevant close to criticality. It increases upon decreasing the colloid-colloid and colloid-substrate distances, and it may reach 25\% of the total force. 
In order to investigate the role of the geometric configuration of the involved bodies for the many-body effects, also the case of three parallel cylindrical colloids was considered\cite{MHD2014}, with the focus on the change of the interaction between two of them due to the presence of a third colloid. Depending on the temperature and the relative positions of the three cylinders, the same interesting features emerged as those mentioned above. This indicates that they are of generic character. 

As in the cases discussed in Sec.~\ref{sec:SF}, a quantitative estimate of {\ccf}s beyond mean-field theory can be obtained via {\mc} simulations of representative models for the universality class of interest. Referring to the case of binary liquid mixtures, the Ising model was simulated in $d=2$,
where the colloids correspond to regions of the lattice with a fixed spin orientation. With a suitably developed cluster algorithm (see Sec.~\ref{ssec:SF-num}) the three-body critical Casimir interaction potential was determined.\cite{HH15} In particular, the case is considered in which two particles (each with the shape of disks) touch each other. The potential is determined as a function of the position of the centre of the third particle. All three particles are of the same size. This highlights the relevance of three-body effects and the large degree of non-additivity of these forces. 
In fact, it turns out that the interaction at sufficiently large distances between the two disks in contact, each of radius $R$, and the third disk is almost identical to the interaction between the third disk and an effective disk of radius $R_{\rm eff}$ in place of the first two.
Interestingly enough, in $d=2$ and at $T=T_c$, this kind of spatial configuration (with two disks at contact) is also 
amenable to an almost  
exact solution. This can be obtained by using conformal invariance and a suitable conformal transformation\cite{HH15}, which confirms the numerical results of the corresponding {\mc} simulations, 
predicting, inter alia, $R_{\rm eff} = \pi R/2$.

Beyond the cases mentioned above, many-body effects have been investigated also in geometrical arrangements which might be relevant for predicting the behaviour of colloidal dispersions under confinement. This setting consists of a pair of spherical colloids which are immersed in a medium confined by two parallel walls forming a narrow slit. This involves a total of four different bodies (two colloids plus two walls)\cite{VDK2018}. In the absence of many-body effects (and thus in the case of pairwise additivity) and if the colloids keep an equal distance from the walls, the component of the force acting on the colloids, which is parallel to the surfaces, would not be affected by the presence of the slit confinement. However, the analysis of this configuration within mean-field theory and {\mc} simulations shows that the effective interactions are strongly non-additive and that the interaction between the two colloids under confinement can be a few times stronger than the corresponding sum of the effective pair potentials. This fact influences, inter alia, the phase behaviour of confined colloids, which acquires a dependence on the slit width.

Many-body effects are expected to be relevant in dense colloidal dispersions or if aggregation occurs --- also due to the critical Casimir interaction. The latter might play a role in determining also the structure of the resulting aggregates or the very phase diagram of the dispersion.  
This was numerically shown to be the case, e.g., in $d=2$\cite{ETB2015}. 
However, beyond what is discussed above and in Sec.~\ref{ssec:exp-mb}, a detailed theoretical or experimental knowledge of the {\ccf}s-induced many-body effects is still lacking. The available experimental data can often be rationalised qualitatively by assuming pairwise additivity of the critical Casimir interactions.  
In particular, various aspects of the phase behaviour of colloidal dispersions in near-critical solvents, which result from the interplay between electrostatic inter-particle repulsion and the pairwise critical Casimir attraction (or resulting from the presence of electrolytes as solvents, see also Sec.~\ref{ssec:FtoC-el}),  were studied experimentally in 
Refs.\!\nocite{BOSGWS2009,VAWPMSW2012,DVBS2013,NFHVS2013,PMVWMSW2013,SNLS2013}\citenum{BOSGWS2009} --\citenum{SNLS2013}, 
and theoretically (analytically or numerically) in 
Refs.\!\nocite{GD2010,MMD2012}\citenum{BOSGWS2009}, \citenum{GD2010} and \citenum{MMD2012}.
%
%
This holds also for a binary mixture of colloids with opposite preferential adsorption\cite{ZAB2011}. 
A recent review of these studies is provided in Ref.~\citenum{MD2018}, so that here we do not discuss them further. 
More recently, in view of possible  applications, the theoretical and experimental studies of colloidal dispersions, which interact also via {\ccf}s, have primarily focussed on the possibility to harness these interactions. This is done as to control the structures of colloids of various kinds resulting from aggregation\cite{MBC-2019} or from the deposition on a substrate\cite{VMKSK2021,MVKSKS2021}:  e.g., quantum dots\cite{MBC-2019}, 
patchy particles\cite{JSSB2021}, or colloidal assemblies in the form of ``superballs''\cite{KSSM2022}.
The phase behaviour and the structure of colloidal dispersions, in which an attractive {\ccf} competes with a repulsive magnetic dipolar interaction (instead of the usual electrostatic or dispersion forces), has also been explored theoretically\cite{MZR2019,MR2020} in $d=2$. 

\subsection{Critical Casimir effect in the presence of electrolytes}
\label{ssec:FtoC-el}

The {\ccf} discussed here, between, say, colloidal particles, is one the various effective interactions which might act in soft matter\cite{I2011}, such as depletion, electrostatic, and van der Waals interactions\cite{P2006}. Differently from these other interactions, which occur rather generically, {\ccf}s emerge as a consequence of the onset of a collective behaviour at critical points. 
Accordingly, they are particularly relevant only in the neighborhood of certain points in the phase diagram of the solvent, although they might be continuously connected with other solvation forces such as depletion forces.\cite{BCPP2010,PBPC2011,GZTS2012}
This is theoretically embodied by the fact that {\ccf}s follow from the contribution $\Omega_V$ (see Eq.~\eqref{eq:Fdec}) to the actual total effective free energy of the confined medium, which is a \emph{non-analytic} function of the thermodynamic distance from the critical point.
The other interactions, instead, contribute to the total effective free energy with terms which are characterised by a smooth dependence on temperature.
Due to their rather different physical origins, the various effective interactions are largely independent of each other, in the sense that they provide additive contributions to the effective interaction between, say, two colloidal particles. 
This is clearly an \emph{approximation}: in fact, e.g., upon approaching the critical point and depending on the effective {\bc}s, adsorption profiles might develop close to the confining surfaces, rendering the confined medium inhomogeneous. This inhomogeneity generically implies a spatial modulation of the frequency-dependent dielectric function $\varepsilon(\omega)$ of the medium at distances close to these surfaces. In turn, this might affect both the possible electrostatic and the van der Waals interactions between them\cite{P2006}, which also depend on $\varepsilon(\omega)$ (see, e.g., the pertinent discussion in Ref.~\citenum{GMHNHBD2009}). 

However, most of the experimental data available so far (see Sec.~\ref{sec:exp}) could be interpreted theoretically by neglecting the interplay between these forces, with the sole exception of the experiment in Ref.\!\nocite{NDHCNVB2011}\citenum{NDHCNVB2011}, which involved a critical binary liquid mixture in the presence of a suitable concentration of electrolytes provided by certain salts.
It was observed that the effective interaction between a charged substrate and a colloid, resulting from the simultaneous action of an expected \emph{repulsive} {\ccf}  and a \emph{repulsive} electrostatic force, might be, depending on the temperature of the medium, \emph{attractive} within a certain range of their surface-to-surface distance $z$. This is shown in, c.f., Fig.~\ref{fig:CvsES}$(a)$, where the interaction potential $\Phi(z)$ is shown as a function of $z$ upon approaching the critical point, i.e., upon increasing the value of $\xi$.
This unexpected observation called for a better theoretical understanding (i.e., beyond the Deby-H\"uckel theory~\cite{I2011}) of the electrostatic interaction occurring between surfaces immersed in a critical solvent and of the subtle coupling between the spatial distribution of the ions in the solvent and the {\op}  of the binary liquid mixture (i.e., the difference in the local concentration of the two components).
Within a simplified lattice model for such a mixture\cite{BGOD2011}, one considers the fact that the space-dependent number densities  $a^{-3}\varphi(\x)$ and $a^{-3}(1-\varphi(\x))$ of the two components of the solvent (taken, for simplicity, to be identical and of volume $a^3$) are generically coupled to the ionic number densities $a^{-3}\varrho_+(\x)$ and $a^{-3}\varrho_-(\x)$ of the cations and anions, respectively. 
In turn, this implies a coupling between the {\op}  $\phi(\x) = \varphi(\x) - \varphi_c$ of the demixing phase transition  of the solvent (occurring at the bulk critical density $\varphi_c$) and the ionic densities, which generates\cite{BGD2012} potentials $V_\pm(\varphi(\x))$ acting on $\varrho_\pm(\x)$. $V_\pm$ accounts for the difference of the bulk solvation free energies of the cations and anions in solvents with $\varphi=1$ and $\varphi=0$, i.e., in the two pure phases of the  mixture (typically water and lutidine). 
This corresponds to the fact that the anions and cations of the salt have different solubilities in the two pure fluid phases formed by the binary liquid mixture. Accordingly, the spatial distributions $\varrho_\pm(\x)$ of the ions are influenced by the spatial variation of the local concentration $\phi(\x)$ of the binary liquid mixture. 
The analysis of the grand potential of this simple, yet rich, model was carried out within mean field theory.
It predicted that the effective interaction potential per unit area $\widetilde\omega(L)$ (see Ref.~\citenum{BGOD2011} for its precise definition) between
two like-charged walls at a distance $L$ and with opposite preferential adsorption can indeed become attractive, 
depending on the ratio $\xi/\kappa^{-1}$ between the correlation length $\xi$ of the critical fluctuations of the order parameter
$\phi$ and the Debye screening length $\kappa^{-1}$ of the electrolyte. This attraction occurs rather generically for $\xi/\kappa^{-1}$ of order 1, while the expected repulsion is found both for $\xi/\kappa^{-1}\gg 1$, which corresponds to the electrostatic repulsion in the almost homogeneous medium, and for $\xi/\kappa^{-1} \gg 1$, in which case the relevant long-distance interaction is provided by the repulsive {\ccf}. 
A plot of the effective interaction potential $\widetilde\omega(L)$, which as a function of $L$ and for various values of $\xi/\kappa^{-1}$ displays the behaviour described above, is provided in, c.f., panel $(b)$ of Fig.~\ref{fig:CvsES}, where it is compared qualitatively with the experimental data reported in panel $(a)$ of the same figure.
The interplay between electrostatic and {\ccf}s described above does not lead to any qualitative changes in the case of critical Casimir attraction and agrees with the experimental observations. 
Similar conclusions were reached by considering various ion-solvent couplings\cite{ST2011,ST2012} and via a Landau-Ginzburg theory such as the one in Eq.~\eqref{eq:LGWwB}, but extended in order to account for the electrostatic phenomena. The Landau-Ginzburg theory leads to a modified Debye-H\"uckel theory for the ion concentration close to a wall and in the presence of critical adsorption\cite{CM2010}. 
This predicts that the interaction between charged walls\cite{PC2011,PCM2012} or colloidal particles (see Ref.\!\nocite{OO2011}\citenum{OO2011} and references therein) changes as described above (see, c.f., the end of Sec.~\ref{ssec:exp-salt}).

The phenomena discussed above are due to the fact that the {\op}  $\phi$ of the phase transition, which is responsible for the emergence of {\ccf} in the presence of confining walls, eventually influences (and is influenced by) other possible interactions within the same system, especially if $\phi$ is spatially inhomogeneous across the film.
Above we considered the interplay between {\ccf}s and electrostatics, motivated by corresponding experimental observations (see also Sec.~\ref{ssec:exp-salt}). However, a spatial variation of $\phi$ is generically accompanied by a variation of the number density of the fluid solvent. This is expected to affect the total contribution of (i) the fluid-fluid interaction and (ii) the fluid-wall interactions to the effective free energy $\Omega_V$ and therefore to the total force. This is particularly the case when these interactions are long-ranged as in the case, e.g., of dispersion forces and therefore there is no clear scale separation between their range, the correlation length $\xi$ of the critical solvent, and the film thickness $L$.
As discussed above for the electrostatic interaction, it turns out that for sufficiently thick films and sufficiently close to criticality, the universal feature of the {\ccf} prevails in the corresponding solvation force. However, away from the critical point, a rich behaviour might emerge also in connection with the occurrence of capillary condensation, investigated in 
Refs.\!\nocite{MDB2004,DSD2007,DRB2007,DRB2009}\citenum{MDB2004} -- \citenum{DRB2009}, 
which the reader is referred to for further details.


\section{Measuring critical Casimir forces}
\label{sec:exp}

In order to have experimental access to {\ccf}s, the relevant fluctuating degrees of freedom have to be able to enter and to leave the space between the confining surfaces.
While this is guaranteed in the case in which at least one of the two involved surfaces is the one of a compact object, e.g., a colloid, this might not always be the case in the film geometry. 
Indeed, the {\ccf}  originates from the free energy change associated with the transfer of degrees of freedom from the space in between the surfaces to the bulk, carried out in such a way that the contribution $-Lp_b$ to Eq.~\reff{eq:Fdec} is the same within the film and in the bulk.  This is not possible if the fluctuating medium is a solid, e.g., a magnet near its Curie point. Indeed, in
this case the transfer of degrees of freedom --- here the magnetic moments ---
would require removing lattice layers from the film which is easy to be
implemented in numerical simulations but difficult to imagine in actual
experiments.
Accordingly,  phenomena such as magnetostriction in films of magnetic materials
and the associated stress should not be considered as a manifestation of {\ccf}s discussed here: the former are indeed due to the variation of the magnetic exchange
interaction as function of the lattice constant (which shrinks as a consequence
of the interaction) and therefore due to a change of the bulk free energy density, which is completely independent of the presence of confining surfaces.
On the other hand, in principle, ferrofluids\cite{GD1994} 
could allow for an observation of {\ccf}s, because, in their fluid phase, the colloids which carry the magnetic moments can enter or leave a confined geometry. However, so far, these systems have not been used in experiments.

Below we briefly review the various experimental investigations of {\ccf}s, focussing particularly on those which allowed their quantitative measurement.

 \subsection{Early indirect evidence}
 \label{subsec:EIE}
 
The occurrence of {\ccf}s among colloidal particles was invoked\cite{HSED98} 
as one of the possible explanations of the early observation of their reversible aggregation\cite{BE85,NKGBGZ1989} in the vicinity of the demixing (critical) point of the solvent.\cite{MD2018} 
However, the available experimental data allowed neither a conclusion in this respect, nor a quantitative determination of the acting force. 
The first attempts to measure \emph{indirectly} the {\ccf}s focussed on the film geometry, which is naturally realised by controllable complete wetting films\cite{KD92b,NI85,I86,Di88} of quantum or classical fluids with a critical point within the fluid phase.  
In fact, if a bulk vapor phase of a fluid, thermodynamically close to the condensation line, is exposed to a suitable substrate at the same temperature, a fluid layer of a certain thickness $L$, known as a \emph{wetting film}, can form on the surface of the substrate. In the presence of strongly attractive substrates, above the so-called wetting transition at vapor-fluid coexistence, complete wetting takes place upon approaching two-phase coexistence\cite{Di88}. Accordingly, $L$ can attain macroscopically large values if, along an isotherm, the undersaturation of the vapor phase vanishes. Under the action of the associated, so-called effective interface potential  \cite{Di88},
the fluid is then naturally confined to the resulting wetting film with the substrate/fluid and the fluid/vapor 
interfaces being perfectly aligned in parallel. For simple fluids, far from $T_c$ the equilibrium thickness $L$ is determined by the competition between the net effect of the dispersion forces between the fluid particles and those of the substrate particles as well as by the undersaturation.
If such a confined fluid is driven thermodynamically towards a continuous phase transition between two distinct liquid phases (but maintaining the phase equilibrium with the common vapor phase),
the associated critical fluctuations at such a critical end point are confined to the film. Accordingly, 
they give rise to a
{\ccf}  $F_\Cas$ as described by Eq.~\reff{eq:FC}. This force acts on the liquid/vapor interface and displaces it from the equilibrium position it exhibits under the influence of the dispersion forces
alone, i.e., in the absence of critical fluctuations.
As a result, the thickness $L$ of the film adjusts its value to a shifted equilibrium position which is caused by the {\ccf}. A detailed study \cite{KD92b} of the dependence of  the thickness of such wetting films on the thermodynamic parameters which control the system (primarily the undersaturation and the temperature) allows an \emph{indirect}  experimental determination of $F_\Cas$ and of the associated scaling function $X_{\rm Cas}$ in Eq.~\reff{eq:FC}.
Following this theoretical suggestion, this kind of experimental approach successfully provided the first quantitative experimental evidence of the occurrence
of {\ccf}s in wetting films ---
of pure liquid $^4$He close its
superfluid phase transition\cite{GC99,GSGC2006} and of a classical
binary liquid mixture\cite{FYP2005} (see, e.g., Ref.~\citenum{ChanChap-23}).


As expected on theoretical grounds --- see Sec.~\ref{sec:SF}, --- in the case of $^4$He
the attractive force leads to a thinning of wetting films.
With classical binary liquid mixtures instead, it is possible to realise both $(+,-)$ and $(+,+)$ effective {\bc}s via a suitable choice of their two components and of the substrate on which wetting occurs. For example, a mixture of methylcyclohexane and perfluoromethylcyclohexane between its vapour phase and a SiO${}_2$ substrate\cite{FYP2005} or, correspondingly, heptane and methanol on a Si  substrate\cite{RBM2007} correspond to $(+,-)$ {\bc}s and lead to a thickening of the wetting film. In the latter case, replacing heptane with nonane turns out to change the {\bc}s from  $(+,-)$  to $(+,+)$.\cite{RBM2007}  
Correspondingly, one observes experimentally\cite{RBM2007} a thinning of the wetting film, which can lead to its collapse if the attractive {\ccf} overwhelms its dispersion counterpart in the effective interface potential. These resulting very thin films no longer probe the universal {\ccf}. In this respect, the observation of Ref.~\citenum{RBM2007} suggests that an attractive {\ccf} is at play but it does not allow  a reliable quantitative determination of such a force, as done in Ref.~\citenum{FYP2005} for the repulsive case.
The qualitative and quantitative features of the scaling functions  $X_{\rm
Cas}$, determined experimentally by studying the wetting films mentioned above, turned out to be in remarkable agreement with the corresponding theoretical predictions, as reviewed, e.g., in Refs.~\citenum{G2009} and \citenum{GD2011}.
%
%
%
\begin{figure*}[t!]
\centering
  \includegraphics[width=0.85\textwidth]{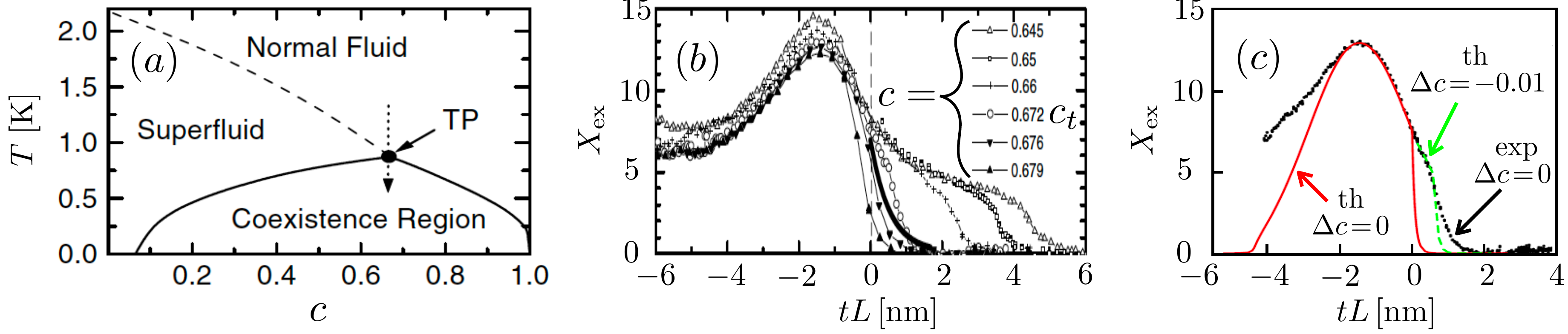}
 \caption{$(a)$ Phase diagram of a $^3$He-$^4$He mixture as a function of temperature $T$ and $^3$He concentration $c$, exhibiting the tricritical point TP. 
 $(b)$ Scaling function $X_{\rm ex}$ of the critical Casimir potential (see Eq.~\eqref{eq:Fdec}) as a function of the (not normalized) scaling variable $t L \, [\mbox{nm}]$, 
with $t = (T-T_t)/T_t$ and for various fixed values of $c$ close to $c_t\approx 0.672$ as determined experimentally\cite{GC2002}. 
 The thick black solid line corresponds to the theoretical prediction for $c=c_t$. It is obtained from a suitable mean-field analysis, which neglects logarithmic corrections, of the extension of Eq.~\eqref{eq:LGWwB} described at the end of Sec.~\ref{sssec-MFT}, with $r=0$ and $u\simeq t$.  $(c)$ The experimental data of panel $(b)$ for $\Delta c= c - c_t =0$ (symbols) are compared with the suitably normalised theoretical predictions (solid red and dashed green lines) which are based on the mean-field analysis of a lattice model (see the main text).
(Panels $(a)$ and $(b)$ are taken from Ref.~\citenum{GC2002}, while panel $(c)$ stems from Ref.~\citenum{MD2006}.)
}
  \label{fig:mix}
\end{figure*}
%
%
We note that the non-universal amplitude $\xi_0^+$, which normalises  the scaling variable $x=(T/T_c-1)(L/\xi_0^+)^{1/\nu}$ of the scaling function  $X_\Cas$, takes different values for the
numerical and the experimental data. In the aforementioned cases, $\xi_0^+$ can be
inferred independently from the temperature dependence of  the corresponding bulk correlation
length $\xi$, which is measured both in the experiments and in the {\mc} simulations. With these normalizations, there are no adjustable parameters left, a fact which makes the agreement  between the theoretical and the
experimental data rather remarkable.

As discussed in the last but one paragraph of Sec.~\ref{sssec-MFT}, at \emph{tricritical points} {\ccf}s arise which belong to a different bulk universality class, such as the point TP occurring in the phase diagram of the $^3$He-$^4$He mixture reported in Fig.~\ref{fig:mix}$(a)$. The corresponding  experimentally controlled thermodynamic variables are the temperature $T$ and the $^3$He concentration $c$ in the mixture, which, at the tricritical point TP, take the values $T_t$ and $c_t$, respectively.
Similarly to the case $c=0$ of pure $^4$He, the scaling function $X_{\rm ex}$ in Eqs.~\eqref{eq:Fdec}  and \eqref{eq:FC} can be determined experimentally  by realising wetting films of thickness $L$ of this mixture close to the corresponding tricritical end point.  
In particular, in Refs.~\citenum{GC2000,GC2000b,GC2002}, $X_{\rm ex}$ was measured  as a function of the scaled deviation $t = (T-T_t)/T_t$ of the temperature from $T_t$ and for various values of $c$ close to $c_t$. 
The resulting repulsive force exhibits a rather rich dependence on the scaling variable $t L$, as shown in Fig.~\ref{fig:mix}$(b)$. This scaling variable naturally emerges from the theoretically expected ones $r L^2$ and $u L$ introduced in Sec.~\ref{sssec-MFT}, taking into account that for $c=c_t$ one has $r=0$ and $u\simeq t$ (see Ref.~\citenum{MGD2007}). We note, however, that the relationship between the pair $(r,u)$ of the parameters of the effective free energy $\HH$ in Eq.~\eqref{eq:LGWwB} and the 
pair of experimentally relevant variables $(t,\Delta c = c -c_t)$ is presently unknown.
(Its determination would require an accurate experimental study of the bulk phase diagram and of the critical properties of the mixture, in particular in the neighbourhood of the tricritical point.)
Accordingly, the comparison between the experimental data and the theoretical predictions is less constrained, and thus less stringent, than in the case of pure $^4$He, in which there is only one relevant scaling variable. Such a comparison is presented in panels $(b)$ and $(c)$ of Fig.~\ref{fig:mix}. In particular, the black solid line for $t\ge 0$ in panel $(b)$ corresponds to the suitably normalised mean-field prediction for $c=c_t$ based on $\HH$, as discussed in Sec.~\ref{sssec-MFT}. Panel $(c)$, instead, compares the experimental data (symbols) for $X_{\rm ex}$ at $\Delta c=0$, extracted from panel $(b)$, with the predictions of the mean-field analysis of the vectoralised Blume-Emery-Griffiths model\cite{MD2006} at $\Delta c= 0$ (red solid line) and at $\Delta c=-0.01$ (green dashed line), after proper normalisation\cite{MD2006,MGD2007}. The discrepancy at $tL \lesssim -20$  is similar to the one observed in the analogous comparison for pure $^4$He, i.e., for $c=0$ (see, e.g., Ref.~\citenum{ZSRKC2007}), while the agreement for $-20\lesssim tL \le 0$ is remarkable. The fact that such an agreement extends to $tL \lesssim 10$ when the comparison is done with the theoretical curve for $\Delta c=-0.01$ seems to suggest that there might be an issue with the experimental determination of $c_t$ in the film geometry. 
A more stringent comparison between the theoretical predictions and the experimental data in Fig. 1(c) would require to express the respective scaling functions in terms of the corresponding (normalised and dimensionless) scaling variable $tL/\xi_0^+$, instead of $tL$ as done in the figure.  Here $\xi_0^+$ is the non-universal amplitude of the singular behaviour of the correlation length upon approaching the tricritical point within the normal phase, according to the general definition introduced in Sec. 2.1.  To our knowledge, the values of $\xi_0^+$ for the mixture as investigated in Refs.~\citenum{GC2000,GC2000b,GC2002} and for the lattice model which provides the theoretical prediction in  Fig.~\ref{fig:mix}$(c)$, are not known. Accordingly, the comparison is presented by normalising the scaling variable of the theoretical prediction in such a way as to fit best the experimental data.
A closer comparison with the experimental data and a determination of the scaling function $X_{\rm ex}$ as a function of two scaling variables certainly stand as interesting open problems. Similarly, it would be interesting to analyse the consequences of the findings of Ref.\!\nocite{IK2022}\citenum{IK2022} concerning wetting near \emph{tricritical points} for the formation of the wetting films considered here.

Additional indirect evidences for the occurrence of {\ccf}s were experimentally sought in the phenomenon of \emph{critical-point wetting} of a binary liquid mixture (see Refs.~\citenum{B99} and \citenum{IB2005} for a discussion). Indeed, the contact
angle $\theta$ formed by the two liquid phases of a binary liquid mixture with a third phase, e.g., a solid, is expected to vanish upon approaching the critical demixing point, realizing complete wetting. However, if a long-ranged attractive critical
Casimir force contributes to the disjoining pressure $\Pi$ (i.e., minus the derivative of the effective interface potential with respect to the film thickness)\cite{UBMCR2003,UBMCFR2003}, which determines $\theta$, this angle may not vanish
at $T_c$ (partial wetting). An accurate determination of the temperature dependence of $\theta$ can provide quantitative information about this force.  Preliminary experiments\cite{UBMCR2003,UBMCFR2003} on wetting of
$^3$He-$^4$He mixtures suggested that the relation $\theta(T=T_c) \neq 0$ was indeed fulfilled (see also Ref.\!\nocite{UB2004}\citenum{UB2004}). 
However, as in the case of the experiment with wetting films, a rather detailed knowledge of the non-critical  background forces is necessary in order to be able to extract reliably a possible contribution of the {\ccf}  from the experimental data. A subsequent experimental study\cite{IB2005} of $\theta$ for the same system unfortunately reached the conclusion that part of the preliminary evidences mentioned above were probably due to optical artefacts, in the course of complete wetting. Furthermore, within the experimental conditions of these studies a possible contribution due to {\ccf}s would be anyhow rather small compared with the contribution due to dispersion forces.
 
The experimental approaches mentioned so far provided first evidences of the occurrence of {\ccf}s in soft matter and, in some cases, they allowed a quantitative determination of such a force.  
However, they also present inherent difficulties associated with their \emph{indirect} nature: in particular, the determination of the {\ccf}  hinges on observing phenomena in which universal and non-universal features are intertwined. These difficulties can be overcome by devising approaches which allow the \emph{direct} measurement of the {\ccf}  or, equivalently, of the associated potential. 
This can be achieved by considering the force acting effectively not on an interface (as in the case of the wetting films mentioned above) but on a spherical particle, which is immersed in the considered fluid and approaches the wall of the container of the critical fluid or another particle. This can be studied via a variety of optical methods (see Ref.~\citenum{CMGV-2021} for a short overview) which will be discussed below.

\subsection{A single colloid close to a homogeneous or a patterned substrate}
\label{ssec:exp-1c}

\begin{figure}[h!]
\centering
  \includegraphics[width=0.35\textwidth]{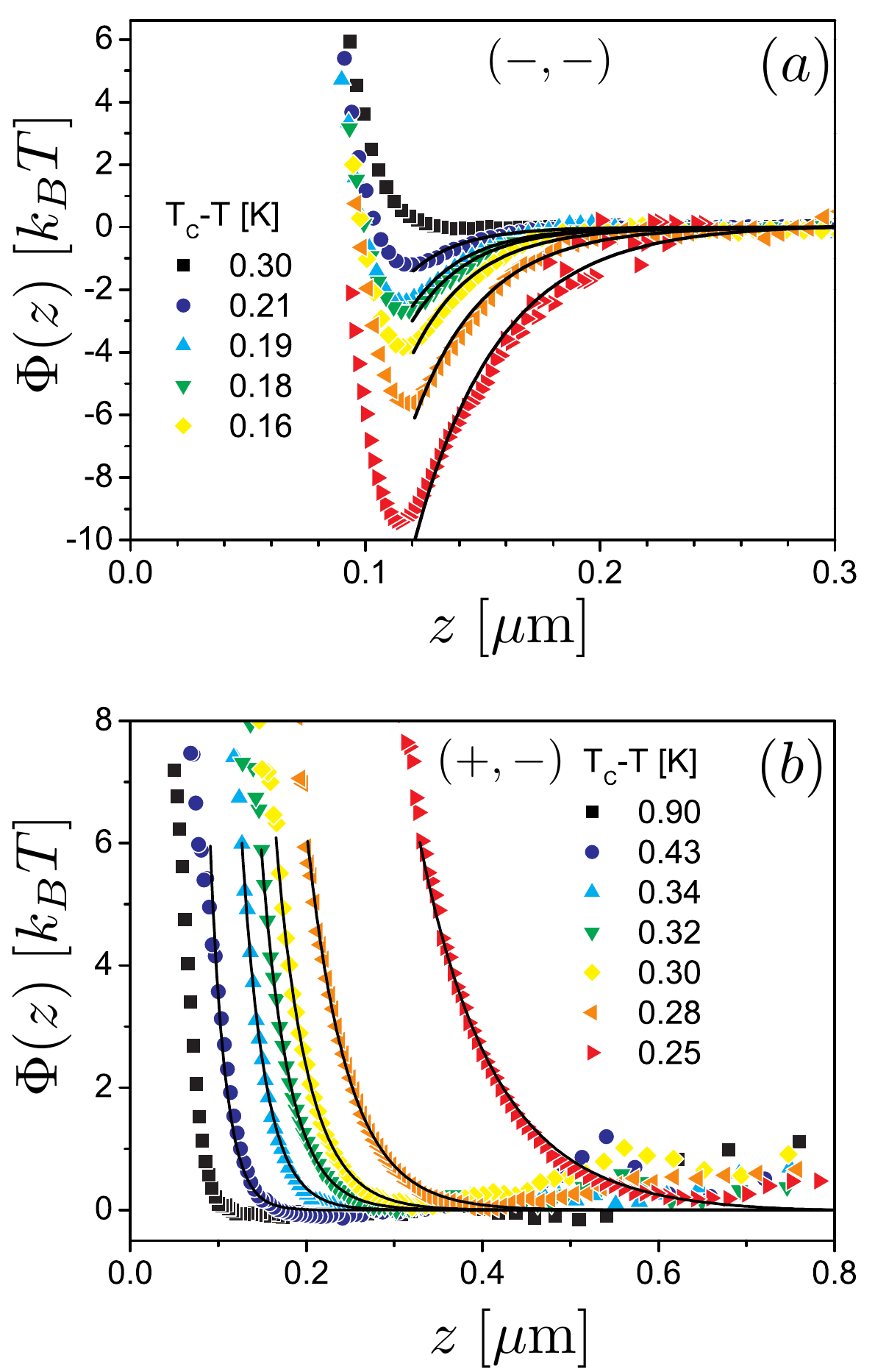}
 \caption{Potential $\Phi(z)$ (symbols) of the forces acting on a colloid immersed in a water-lutidine mixture and at surface-to-surfaces distance $z$ from an hydrophilic wall,  for various temperatures  $T$ of the mixture at the critical concentration~\cite{HHGDB2008,GMHNHBD2009}. Upon increasing $T$ towards the critical
  value $T_c\approx 34^\circ\mbox{C}$, an attractive {\ccf}  is observed with a hydrophilic
particle [$(-,-)$ {\bc}s, panel $(a)$], whereas with a hydrophobic particle [$(+,-)$
{\bc}s, panel $(b)$] the force is repulsive. 
The solid lines in $(a)$ and $(b)$
correspond to the theoretical predictions for the contribution $\Phi_\Cas(z)$ [see Eq.~\eqref{eq:ccpot-da}] of
the {\ccf}  to the total potential $\Phi(z)$. The value $\nu \approx 0.63$ of the critical exponent of the correlation length $\xi(T)$ is fixed by the Ising universality class, while the non-universal amplitude $\xi_0^+\approx 0.2\,\mbox{nm}$ was independently determined by light scattering experiments for this mixture (see Refs.~\citenum{HHGDB2008} and \citenum{GMHNHBD2009} for further details). Accordingly, there are no adjustable parameters.
(Plots taken from Ref.~\citenum{HHGDB2008}.)
}
  \label{fig:direct}
\end{figure}

In the case of a spherical particle of radius $R$ approaching a flat surface, both the wall and the sphere impose effective {\bc} onto the {\op}  $\phi$. Therefore, as discussed in Sec.~\ref{sec:FtoC}, a {\ccf}  $F_C$ acts on the sphere. If the distance  $z$ of closest approach of the sphere from the homogeneous wall is much smaller than $R$, the potential $\Phi_\Cas(z)$ of the {\ccf}  takes the form given in Eq.~\eqref{eq:ccpot-da}.
Within the Derjaguin approximation, the function $\vartheta_{|\circ}(y)$ can be expressed in terms of 
$X_\Cas(y)$, as discussed in Sec.~\ref{ssec:FtoC-DJ}.
These functions 
share their qualitative and universal properties. 
Within this geometrical setting, the experimental challenge is to
detect the onset of the contribution $F_\Cas$ to the total force acting on the sphere
as the critical point of the surrounding fluid is approached. In order for the
effects due to $F_\Cas$ to be detectable for such a sphere floating in the mixture,
$F_\Cas$ has to be comparable in magnitude with the typical forces at play. This
suggests the use of colloids, i.e., micrometer-sized particles, dispersed in
classical binary liquid mixtures. Indeed, the typical energy scale of the
interaction within this system is provided by $k_{\rm B}T$, i.e., the same scale
as the one for $\Phi_\Cas(z)$ in Eq.~\reff{eq:ccpot-da}.
The necessary sensitivity for the measurement of the force could be achieved by using 
total internal reflection microscopy\cite{P99,CMGV-2021,D3SM00085K} (TIRM),  which is actually capable to measure forces with femto-Newton force resolution.
With this technique it was possible\cite{HHGDB2008,GMHNHBD2009} to study the onset of {\ccf}s acting on a single colloid near a wall as the temperature $T$ of a water-lutidine mixture, used as the solvent, is increased towards its (lower) critical value $T_c$ 
at fixed lutidine critical mass fraction. 
Figure~\ref{fig:direct} shows the experimental results and their comparison with the available theoretical predictions.

With reference to the water-lutidine mixture, ``$+$'' and ``$-$'' indicate the preferred adsorption of lutidine and water, respectively. Accordingly, a hydrophilic colloid (with $R\approx 1\,\mu\mbox{m}$) close to a hydrophilic wall realises $(-,-)$ {\bc}s. In this case, the measured potential $\Phi(z)$ in Fig.~\ref{fig:direct}(a) for $T_c-T=0.3\,\mbox{K}$, i.e.,
sufficiently far from the critical point, results only from the combined action of buoyancy and of the electrostatic repulsion between the colloid and the wall. (In fact, as discussed in Ref.~\citenum{GMHNHBD2009}, dispersion forces turn out to be negligible in this experiment.)
This repulsion is due to the fact that surfaces in contact with highly polar fluids acquire a surface charge after releasing a negligible amount of ions into the mixture. Buoyancy, instead,
is characterised by a well-known linearly increasing potential which has been subtracted from all the data
presented in Fig.~\ref{fig:direct}. 
Upon approaching the critical point, an increasingly deep potential well develops as a consequence of the
attractive {\ccf} which provides a negative contribution $\Phi_\Cas(z)$
to the potential of the remaining forces. As expected theoretically, the spatial range $\xi$ of $\Phi_\Cas(z)$ increases
for $T\to T_c$. The set of measured potentials $\Phi(z)$ can be
compared with the theoretical prediction in Eq.~\eqref{eq:ccpot-da} for  $\Phi_\Cas(z)$ ---
within the range of distances $z$ for which the
(non-critical) electrostatic interaction (basically given by the curve
corresponding to $T_c-T=0.3\,\mbox{K}$) is expected to be negligible. Within
this range, the theoretical predictions are reported as solid lines in
Fig.~\ref{fig:direct}(a) and indeed they are in excellent agreement with the
experimental data.
%
%
In the case of a hydrophobic colloid (with $R\approx 2\,\mu\mbox{m}$), the corresponding {\bc}s change from $(-,-)$ to $(-,+)$ and, on theoretical grounds, repulsion is expected. Indeed, far below the
critical point, the potential $\Phi$ in Fig.~\ref{fig:direct}(b) for $T_c-T=0.9\,\mbox{K}$ consists  of the electrostatic
repulsion only. Upon approaching the critical point, the repulsive part of the potential curves shifts towards
larger values of $z$ due to the fact that, as expected, increasingly strong repulsive Casimir forces
act on the particle and yield a positive contribution $\Phi_\Cas(z)$ to the
total potential $\Phi(z)$.
The corresponding theoretical predictions in Eq.~\eqref{eq:ccpot-da} for  $\Phi_\Cas(z)$ are reported as solid lines in {\mc}Fig.~\ref{fig:direct}(b). Within the region where the
non-critical contribution to the interaction is negligible the agreement with
the experimental data for $\Phi(z)$ is remarkable.
If, via a suitable chemical surface treatment, the adsorption preference
of  the silica wall is changed from being hydrophilic $(-)$ to hydrophobic $(+)$, attraction is recovered, in agreement with the theoretical expectations and predictions\cite{GMHNHBD2009}.
In contrast to the smooth and reversible onset of the {\ccf} , in mixtures with 
a lutidine mass fraction $x$ smaller than the critical one $x_c$ and with $(+,+)$ {\op} or with 
$x>x_c$ and $(-,-)$ {\bc}s, one observes the abrupt formation of a potential well upon approaching the phase transition line\cite{HHGDB2008,GMHNHBD2009}. This fact can be interpreted in terms of the onset of liquid bridge formation between the particle and the wall. This occurs only if the mixture is poor in the component preferred by both the colloid and the wall. This tells that surface contributions to the free energy become
sufficiently important so that the system characterises its interfacial energy via bridge formation.

The \emph{universal} character of the {\ccf}  was confirmed, at least in the attractive case, by 
experiments\cite{HKTAGB2021} which used an aqueous critical micellar solution of the nonionic surfactant $\mbox{C}_{12}\mbox{E}_{5}$,  instead of the standard molecular binary liquid mixture of water and lutidine primarily considered in experiments. Different from the latter, at the concentrations explored in the experiment, the surfactant forms wormlike micelles which are ca.~10 times larger than the molecules involved in the molecular liquid. 
In spite of the rather different microscopic structure of these two solvents, it turned out\cite{HKTAGB2021} that their confinement gives rise to the same  {\ccf}s 
on a colloid opposing a substrate or on two opposing colloids, both in agreement with the theoretical predictions.
Differences between the interactions measured for these two solvents emerge only at small inter-particle distances, where other, possibly different, non-universal contributions come into play.

The experiments briefly presented above demonstrated that the {\ccf}s can be strong enough to influence significantly the behaviour of soft matter at the sub-micrometer scale. Compared with the various interactions typically acting in these systems --- such as electrostatic, dispersion, entropic ones etc. ---  the {\ccf}  is characterised by an uncommonly high degree of tunability with respect to strength, range, and sign. All these features can be harnessed in order to serve dedicated purposes such as for colloidal suspensions.
In order to explore these possibilities and to have  full experimental control of {\ccf}s, a variety of systems involving colloidal particles has been investigated by means of various experimental techniques. They range from light scattering to digital video microscopy and surface plasmon spectroscopy.\\[-3mm]  

\emph{Inhomogeneous surfaces.---} In particular, chemically patterned substrates can be used to produce a lateral modulation of the {\ccf}. This allows one to control \emph{reversibly} the spatial distribution and the formation of structures in a dilute colloidal suspension with a near-critical solvent\cite{SZHHB2007}.
These experimental studies, together with the corresponding theoretical analyses\cite{TKGHD2009,TKGHD2010}, have revealed, inter alia, that the sensitivity of the resulting force with respect to details of the imprinted patterns can even be exploited in order to infer their fine geometrical features\cite{TZGVHBD2011}. 
For example, from monitoring the spatial distribution of colloids above the surface of a substrate containing  a chemical stripe (realised, e.g., by removing an atomic layer from a surface coating), it is possible to distinguish stripes with a well-defined width from those the widths of which fluctuate due to the fabrication technique (using focussed ion beam vs.~microcontact printing, see  Ref.~\citenum{TZGVHBD2011}).
Suitable surface treatments allow one to realise substrates which are characterised by preferential adsorption for one of the components of a binary liquid mixture which varies gradually along a certain direction\cite{NHC2009}. 
These substrates have been used for investigating the {\ccf} in the 
presence of variable boundary  conditions, giving rise to crossover phenomena. 
These experimental results, reviewed in Ref.~\citenum{GD2011}, offer possible applications towards controlling actively the rheology and diffusion of colloidal suspensions which are exposed to such, engineered substrates.

\subsection{More than two colloids and many-body effects}
\label{ssec:exp-mb}

The theoretical predictions discussed in Sec.~\ref{ssec:FtoC-mb} showed that the expected many-body effects in the critical Casimir interaction exhibit a rather complex dependence on the relative positions of the involved bodies, their geometrical shape, and the temperature of the medium; moreover they might be quantitatively rather significant. 
However, it has been unclear whether they could be detected in actual experiments involving colloids. This issue was addressed in Ref.\!\nocite{PCT-2016}\citenum{PCT-2016} by studying the interaction between two trapped colloidal particles in the presence or in the absence of a third particle close to them, upon approaching the (lower) critical point of the water-lutidine mixture. The spatial positions of these three fluctuating colloids were controlled via holographic optical tweezers\cite{CMGV-2021}. 
These positions 
(actually projected onto the horizontal plane identified by the beam waists of the three optical tweezers, centred at the vertices of an equilateral triangle with the edge $\approx 2.3\mu{\rm m}$ long)
were measured via digital videomicroscopy, such as to determine the statistical distribution of the thermally fluctuating, inter-particle 
projected distance $l_{12}$ between two particles (labeled 1 and 2) forming one of three possible pairs.
%
%
\begin{figure}[h!]
\centering
  \includegraphics[width=0.37\textwidth]{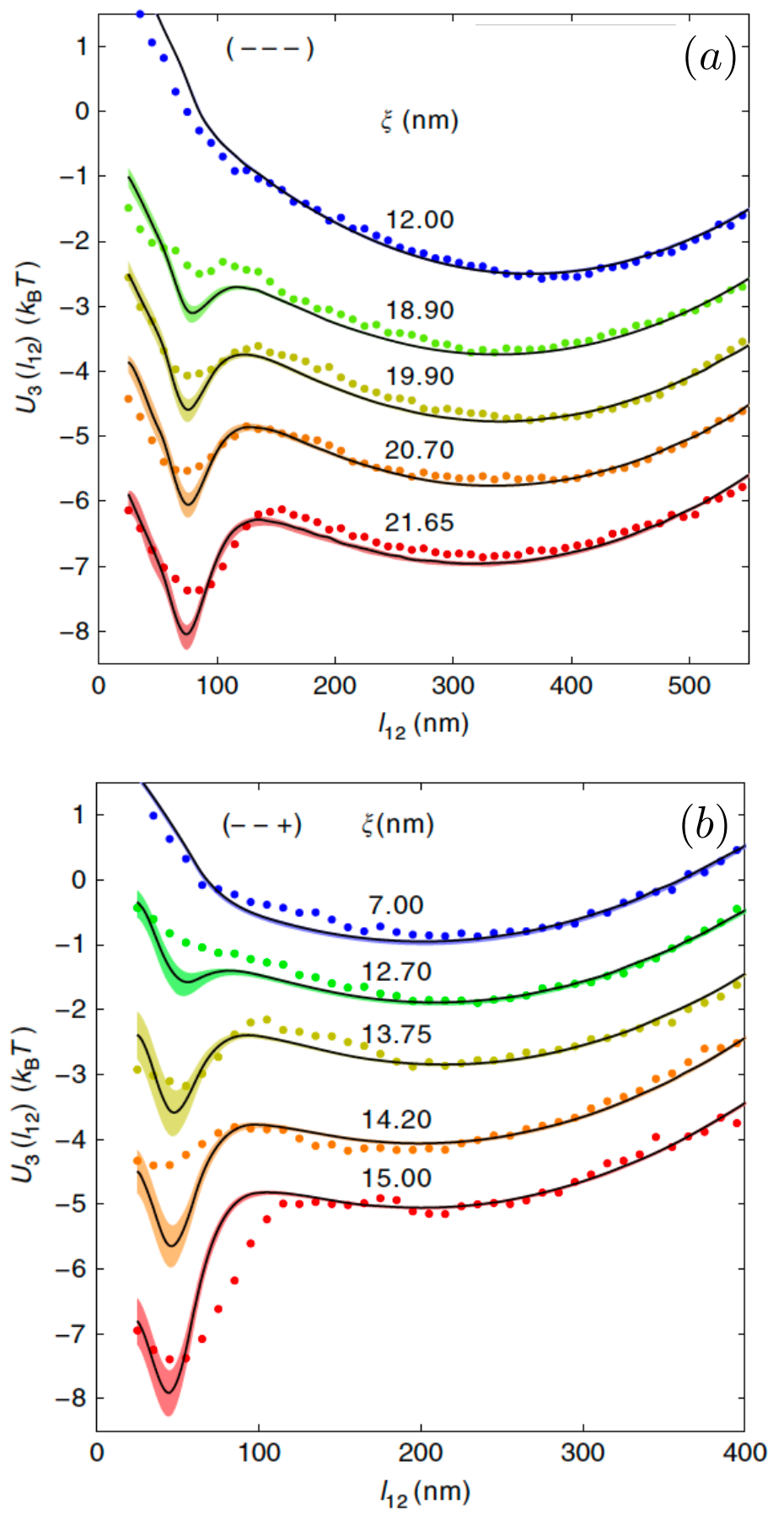}
 \caption{Effective potential $U_3(l_{12})$ (symbols) obtained from the distribution of the projected distance $l_{12}$ between colloidal particles number 1 and 2, in the presence of the colloidal particle number 3 with $(a)$ the same or $(b)$ the opposite  adsorption preference as the first two hydrophilic  ($(-)$ \bc) colloidal particles. The solid lines correspond to  the effective potential expected by assuming pairwise additivity of the interaction between the three pairs of particles (see the  main text).  The shading associated with each solid line quantifies the systematic uncertainties in the determination of these pair potentials. Upon increasing the correlation length $\xi$, i.e., for $T\to T_c$,  the emerging discrepancy between the solid lines and the symbols indicate the presence of many-body effects. For the ease of visualisation, the subsequent curves in panels $(a)$ and $(b)$ have been separated vertically by an inconsequential constant shift of 1 $k_{\rm B}T$. 
 All colloidal particles have a radius $R \approx 1 \mu\mbox{m}$.
 (The plots are taken from Ref.~\citenum{PCT-2016}.)
}
\label{fig:mb}
\end{figure}
%
In the absence of the third particle (i.e., if it is kept sufficiently far from the other two), this distribution was found to be well fitted by that one expected on the basis of the theoretical predictions for the inter-particle critical Casimir potential (within the Derjaguin approximation, see Sec.~\ref{ssec:FtoC-DJ}) and for the electrostatic repulsion. This comparison allows one to determine the material parameters of the system and therefore to characterise, e.g., the electrostatic repulsion as well as the stiffness of the optical trapping. When the third particle is brought close to the previous two, it affects the distribution $P_3(l_{12})$ of the  inter-particle, projected distance $l_{12}$ between the first two particles, with the corresponding effective potential $U_3(l_{12}) = - k_{\rm B} T\ln P_3(l_{12})$. This experimentally determined effective potential, indicated by the symbols in Fig.~\ref{fig:mb}, was compared with the one expected  due to assuming pairwise-additivity of the interaction between the three pairs of particles and by using the pair potentials  determined previously. This expected potential is indicated by the solid lines in Fig.~\ref{fig:mb}, where the shading indicates the systematic uncertainties in the determination of the pair potentials.
Discrepancies between symbols and solid lines are observed beyond the statistical and systematic uncertainties, especially upon increasing the correlation length $\xi$, i.e., upon approaching the critical point. These discrepancies are observed independently of the fact whether the third particle has the same adsorption preference as the first two (realising $(---)$ {\bc}s for the three colloids) or not (realising $(--+)$ {\bc}s), as shown in panels $(a)$ and $(b)$ of Fig.~\ref{fig:mb}.
This demonstrated the occurrence of genuine many-body effects. This is particularly the case within the range of distances $l_{12}$ at which {\ccf}s provide the dominant contribution to the effective interaction --- compared with, say, the electrostatic repulsion. This contribution causes, upon decreasing $l_{12}$, the sharp drop into the dip displayed by $U_3(l_{12})$ in Fig.~\ref{fig:mb} at short distances.

As anticipated in Sec.~\ref{ssec:FtoC-mb}, besides many-body effects, the influence of the {\ccf}  on the physical properties of colloidal suspensions is significant. This is confirmed by a number of theoretical and experimental studies briefly mentioned in that section and reviewed in Ref.~\citenum{MD2018}, which the reader is referred to for further details. 
Here we only mention that upon changing the temperature-dependent range of the critical Casimir interaction between two 
colloidal particles in a near-critical solvent, it is possible to induce and access experimentally a liquid-gas phase separation of the colloidal fluid and possibly the associated critical point. 
This was observed for solvents consisting of a molecular binary liquid mixture\cite{ZAB2011,NFHVS2013} or of
an aqueous solution of globular solid micelles\cite{BCPP2010,PBPC2011}.
Moreover, as a convenient way to engineer and control the total interaction energy  between colloids in suspension, the {\ccf}  can be used in order to study static and dynamic phenomena. 
They depend on the features of this interaction, such as the formation of structures resulting from aggregation\cite{VAWPMSW2012,PMVWMSW2013}, which may change in zero gravity.

\subsection{Interplay with other forces}
\label{ssec:exp-salt}

In Sec.~\ref{ssec:FtoC-el}  we anticipated  that {\ccf}s in soft matter usually act together with forces of different nature such as electrostatic, van der Waals, or depletion forces. Accordingly, it is natural to explore the extent to which their corresponding effects simply add or if these various forces might influence each other. In most of the cases of experimental relevance and in view of the currently available experimental accuracy it turns out that this possible (and somehow expected) mutual influence is quantitatively negligible within the range of parameters studied so far. A notable exception was provided by the investigations of {\ccf}s emerging in binary liquid mixtures with added salt\cite{BOSGWS2009,NDHCNVB2011}. In fact, in a sufficiently dilute colloidal suspension, one expects that the only relevant interactions are provided by the electrostatic repulsion and the critical Casimir attraction, the spatial ranges of which are controlled by the Debye screening length $\kappa^{-1}$ and the correlation length $\xi$, respectively. Depending on the ratio of these two lengths and thus on the relationship between temperature and ion concentration in solution, one expects  the resulting total potential to be capable to drive aggregation or not. The investigation of this phenomenon in the $\xi$--$\kappa^{-1}$ plane\cite{BOSGWS2009} (see also Refs.~\citenum{GD2010} and \!\nocite{BWS2010}\citenum{BWS2010}) motivated a better understanding of the interaction between a (charged) colloid and a surface in the presence of an electrolyte.  
%
%
\begin{figure}[h!]
\centering
  \includegraphics[width=0.37\textwidth]{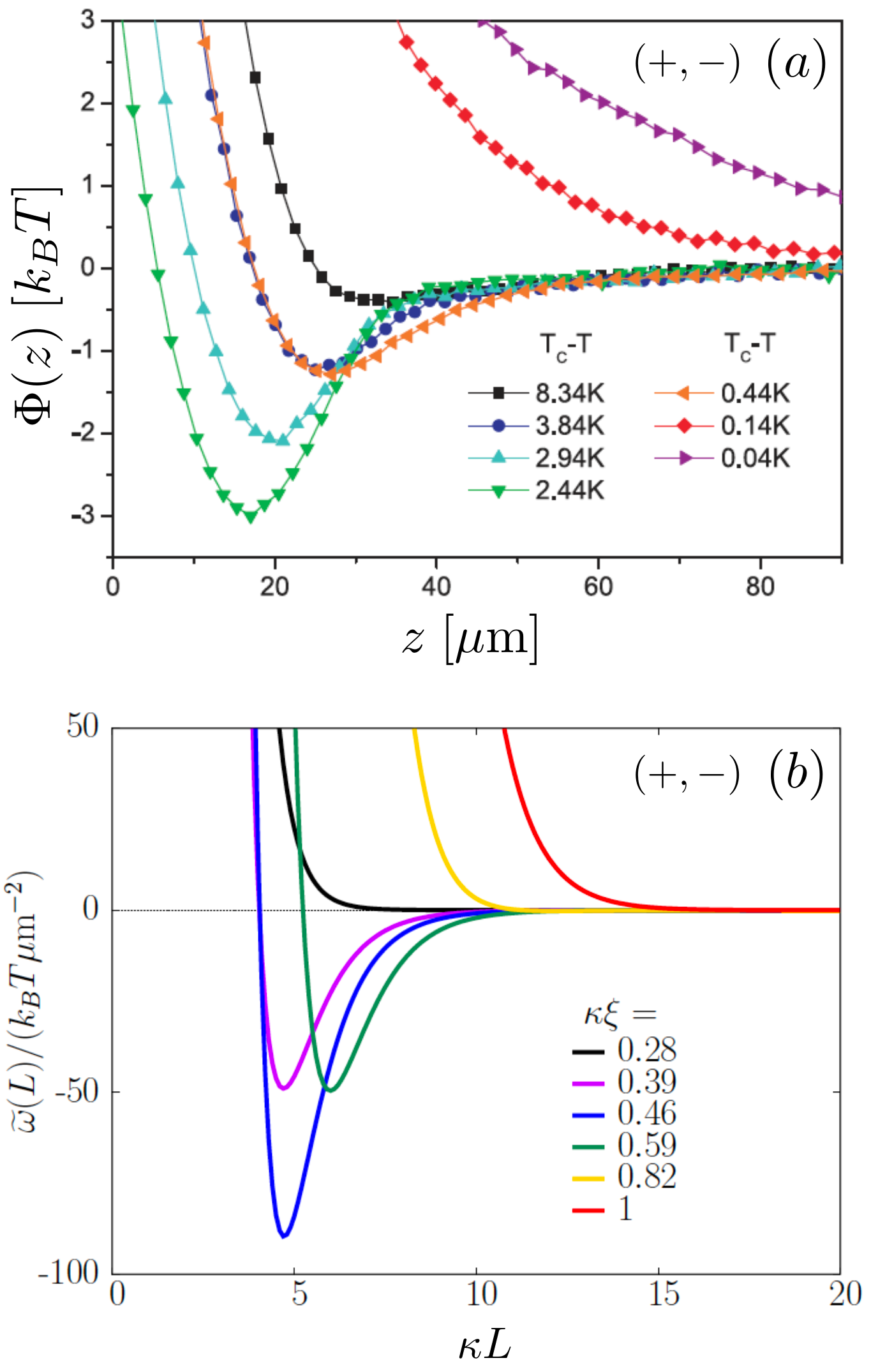}
 \caption{Effective interaction between $(a)$ a colloidal particle at a surface-to-surface distance $z$ from a homogeneous substrate or $(b)$ two parallel plates separated by a distance $L$, immersed in a critical binary liquid mixture in the presence of 
 electrolytes provided by certain salts. The two surfaces involved have opposite adsorption preferences, as to realise $(+,-)$ {\bc}s, giving rise to a \emph{repulsive} {\ccf}. The salt concentration controls the Debye screening length $\kappa^{-1}$, while the deviation of the temperature $T$ from its critical value $T_c$ determines the correlation length $\xi$. 
Panel $(a)$ reports the experimental data for the potential $\Phi(z)$ obtained\cite{NDHCNVB2011} via TIRM with a (hydrophilic) polystyrene colloid (realising $(-)$ {\bc}s) of radius $R\approx 0.75 \mu{\rm m}$ in a mixture of water and 2,6-lutidine containing 10 mM KBr. 
Panel $(b)$ shows the effective interaction potential $\widetilde\omega(L)$ as a function of $L$ and for various values of $\kappa \xi$, as obtained theoretically within the model  discussed briefly in Sec.~\ref{ssec:FtoC-el}. Although the theoretical predictions displayed here depend quantitatively on the choice of the values of various parameters involved in such a model (see Ref.~\citenum{BGOD2011} for details).  Their qualitative features and the dependence on $\kappa \xi$ are rather robust and agree with the experimental observation shown in panel $(a)$.
(Panels $(a)$ and $(b)$ are taken from Refs.~\citenum{NDHCNVB2011} and \citenum{BGOD2011}, respectively.)
}
\label{fig:CvsES}
\end{figure}
%
The corresponding potential $\Phi(z)$ was experimentally determined via TIRM (see Sec.~\ref{ssec:exp-1c} and Ref.~\citenum{NDHCNVB2011}). 
As anticipated in Sec.~\ref{ssec:FtoC-el}, the attractive force, observed for \emph{similar} adsorption preferences at the colloid and the surface, i.e., for $(\pm,\pm)$ {\bc}s, was qualitatively not affected by the presence of the salt. The phenomenon observed in the presence of \emph{opposite} adsorption preferences, i.e., for $(\pm,\mp)$ {\bc}s (leading, in the absence of salt, to a repulsive {\ccf}) was richer, as shown in Fig.~\ref{fig:CvsES}$(a)$. 
In fact, starting at a temperature $T$ far from $T_c$, first one observes  the expected electrostatic repulsion (as in Fig.~\ref{fig:direct}$(b)$). Then, upon approaching $T_c$, an \emph{attractive} contribution to the total potential develops before this gradually fades and a predominantly \emph{repulsive} component eventually emerges close to criticality, as expected in the absence of salt and in line with the concept that asymptotically for $T \to T_c$ the \emph{universal} scaling function should be recovered.
This experimental observation motivated the theoretical investigations discussed in Sec.~\ref{ssec:FtoC-el}, which eventually turned out to explain qualitatively the phenomenon, as shown in Fig.~\ref{fig:CvsES}$(b)$.
A quantitative analysis, however, is still lacking, as it would require a careful characterization of the material properties of the system (such as the surface charges, etc.) which crucially determine the non-universal interactions within this system.


\section{Dynamics}
\label{sec:Dyn}

The previous sections focussed on the \emph{stationary} properties of {\ccf}s occurring \emph{in thermal equilibrium}. However, these forces are expected to manifest themselves also in \emph{non-stationary} and possibly in \emph{out-of-equilibrium} conditions, such as those generated by sudden perturbations of the system due to temperature changes,\cite{G2008,RKK2017,GRD2019} the action of external forces\cite{GaD2006}, 
or due to the non-equilibrium nature of the stationary state in which the confined fluctuations occur\cite{CBMNS2006,BMS2007,FGG2021}.
Considering non-equilibrium conditions or time-dependent properties in equilibrium 
(such as temporal correlation functions of the forces) poses a conceptual challenge to the approaches discussed in Sec.~\ref{sec:SF}. In fact, while the equilibrium expectation value of the force can be derived from the effective free energy $\Omega$ or, alternatively, from the expectation value of the stress tensor $T_{ij}$, only the latter option is viable out of equilibrium\cite{G2008}, taking into account dependence of the {\op}  field $\phi(\x,t)$ on time $t$. (Note, however, that the necessity of this choice has been debated.\cite{DG-2009,DG-2010,KSD-2018}) 
Accordingly, the dynamical behaviour of {\ccf}s can be predicted on the basis of the dynamics of the {\op}  field $\phi(\x,t)$. For the sake of brevity, here we shall not present in detail the well-established theory of dynamical critical behaviour\cite{HH77,Tau-2014}.
Instead, we provide the theoretical prediction and the experimental evidence that universality emerges also in this case, in the sense that, on the long time scale, the collective dynamical behaviour of the system is determined by the gross features of its dynamics, such as symmetries and possible conservation laws, and is largely independent of the microscopic details of the dynamics. 
In analogy to what occurs in equilibrium, this is due to the fact that also the relevant timescale $\tau$ of the dynamics displays an algebraic singularity as the critical point is approached, i.e., $\tau\sim \xi^z$, where $z$ is the so-called \emph{dynamic critical exponent}, which characterises the dynamic universality class. 
In particular, from the theoretical point of view, the dynamics of the {\op}  field is generically expressed in terms of stochastic evolution equations with thermal white noise, possibly involving additional fields beyond the {\op}  itself. 
If the system under investigation is in thermal equilibrium, these stochastic differential equations satisfy the fluctuation-dissipation theorem, and the equilibrium universality class might split into various dynamic universality classes, depending on possible conservation laws. 

\subsection{Analytical and numerical studies }
\label{ssec:Dyn-th}

Among the simplest possible instances of \emph{equilibrium} dynamics for a scalar {\op}  field $\phi(\x,t)$ are the so-called models A and B 
of Ref.\!\nocite{HH77}\citenum{HH77}, according to which 
\begin{equation}
\partial_t \phi(\x,t) = -D (-\nabla^2)^a \frac{\delta \HH 
}{\delta\phi(\x,t)} + \zeta(\x,t).
\end{equation}
Here $\HH$ is the Landau-Ginzburg effective free energy given in Eq.~\eqref{eq:LGWwB} (omitting, in general, the boundary terms),
$\zeta(\x,t)$ is a Gaussian white noise with $\langle \zeta(\x,t) \rangle =0$ and correlation function  $\langle \zeta(\x,t) \zeta(\x',t') \rangle = 2 D k_{\rm B}T (-\nabla^2)^a\delta(\x-\x')\delta(t-t')$, where $a=0$ corresponds to model A and $a=1$ to model B. In the latter case, the {\op}  $\phi$ is locally conserved in the sense that $\partial_t \phi = - \bm{ \nabla}\cdot \bm{J}$ for a suitable stochastic current $\bm{J}$, as it occurs if $\phi$ is the {\op}   of a binary liquid mixture. However, we note that the actual dynamics of these mixtures is significantly more complicate, because one has to account also for the coupling of $\phi$ to other slow modes, such as the fluid velocity.
This is encoded in the so-called model H of Ref.~\citenum{HH77}, 
briefly discussed further below.

As in the case of the stationary properties, the dynamics is affected by the presence of boundaries (see, e.g., Ref.~\citenum{D86}). The stochastic evolution of $\phi(\x,t)$ (and possibly additional conserved quantities) has to be supplemented by the {\bc}s discussed in Sec.~\ref{sec:CP-U}. This possibly leads to a further splitting of the dynamic universality class depending on the presence or absence of locally conserved dynamics at the boundaries.\cite{D1997}
Similarly, the dynamics in the presence of spatial confinement can be theoretically investigated by considering the simultaneous presence of various boundaries. However, up to now, analytical progress has been possible only for the aforementioned simplest cases of dynamics and for the film geometry. In this respect, we note that the dynamics of {\ccf}s is much less explored than their stationary properties in spite of the fact that these dynamical properties are within experimental reach and might be investigated in the near future. \\[-3mm]

\emph{Film geometry.---} In particular, the dynamics of a scalar Gaussian field $\phi(\x,t)$  (corresponding to $u=0$ in Eq.~\eqref{eq:LGWwB}), confined within a film with ordinary {\bc}s) was considered in Ref.\!\nocite{GaD2006}\citenum{GaD2006}. If $\phi(\x,t)$ is locally perturbed inside the film, the resulting critical Casimir pressures, acting on the two confining surfaces, become time- and space-dependent. These two pressures are no longer necessarily equal as they are in thermal equilibrium.  Within model A, the associated scaling function was determined after switching off the perturbation. This analysis was extended\cite{G2008} to the case of different {\bc}s at the two confining surfaces (Dirichlet/Neumann), and for a global quench of the temperature of the medium, i.e., for a sudden change from a very high value to the critical point. In both cases, the strength of the force on the two surfaces turns out to be different while, as expected, it approaches the same value as the relaxation towards an equilibrium condition occurs. 
In an analogous setting, the case of a Gaussian field with short-ranged spatial correlations (modelled by neglecting the term $\propto (\nabla\phi)^2$ in Eq.~\eqref{eq:LGWwB}), conserved dynamics (model B), and no-flux condition $\bm{J} = \bm{0}$ at the boundaries was considered in Ref.~\citenum{RKK2017}.  In particular, upon quenching the strength of the noise (i.e., the temperature) from an initial state with $\phi=0$ (and therefore a vanishing fluctuation-induced force), the emergence of a transient, long-ranged dynamical effective Casimir-like force also away from criticality was observed. 
This is the case because the emergence of these long-ranged correlations is due to the conservation law of the dynamics. However, this force disappears in the stationary state, which is attained long after the quench, 
because the correlation length of the fluctuations within the medium was assumed to be vanishingly small.
The case of inclusions embedded in the fluctuating field and the possible realization of the aforementioned  quench in active matter (see also below) were discussed. 
The interplay between the long-ranged correlations generated out of equilibrium, in accordance with model B, and those emerging when the system is poised at its critical point, becomes particularly rich after a quench to this point. In Ref.~\citenum{GRD2019},
the resulting non-equilibrium critical Casimir effect in the film geometry was investigated for a Gaussian field with non-symmetry-breaking {\bc}s. In particular, it turns out that complex transient regimes arise, including correlation functions and forces with oscillatory evolution, depending on the conditions in which the medium is prepared ahead of the quench to the critical point and depending on the specific choice of the {\bc}s. 

Alternatively, a non-equilibrium condition can be generated by modulating externally, as a function of time $t$, the distance $L(t)$ between a mobile plate and a fixed one. The two plates confine a scalar, fluctuating medium $\phi$ to a film with Dirichlet {\bc}s. 
As a consequence, the force acting on the fixed plate is also modulated and its dynamics and universal form were determined analytically for a critical, scalar Gaussian field with model A dynamics and for a  harmonic dependence of $L(t)$ on $t$, under the assumption of a small modulation amplitude compared to the average distance between the boundaries.\cite{HA2013} In this limit, the resulting force also depends harmonically on time, with the same angular frequency $\omega_0$ as $L(t)$ does, and with a $\omega_0$-dependent amplitude and phase shift. In addition, this force displays two distinct contributions, one generated by diffusion of stress within the medium and another one is related to resonant dissipation in the cavity (see Ref.~\citenum{HA2013} for details).

The theoretical investigations mentioned above considered dynamical evolutions in a confined geometry with {\bc}s which do not break the $\phi \leftrightarrow -\phi$ symmetry of the Gaussian model. The choice of the latter properties are primarily motivated by the search for analytical solutions. 
However, as extensively discussed in the previous sections, in many cases of experimental interest the boundaries are responsible for critical adsorption, which breaks such a symmetry and can be effectively represented by introducing surface fields, such as in Eq.~\eqref{eq:LGWwB}. As a consequence of this choice, 
$\langle \phi(\x,t)\rangle$ exhibits a non-trivial evolution, which can be conveniently studied within a mean-field approximation of the evolution equations. 
In Ref.\!\nocite{GGD2018}\citenum{GGD2018}, the case of a scalar {\op}  was studied with the effective free energy $\HH$ given by Eq.~\eqref{eq:LGWwB}, with model B dynamics, and with no-flux {\bc}s at the $(+,+)$ boundaries of a film. The dynamics of the film was considered after an instantaneous quench from a homogeneous, high-temperature state with $\langle \phi(\x,t)\rangle = 0$ to $T_c$  or to still supercritical but lower temperatures. 
The evolution of the {\ccf}  acting on the confining surfaces was determined together with the corresponding dynamic scaling functions. 
It turns out that such a force contains a non-equilibrium contribution which, depending on the parameters of the system, might have a non-monotonic dependence on time, while being generically \emph{repulsive}. It actually turns out\cite{GVGD2016} that even the attractive or repulsive character of the {\ccf} in \emph{equilibrium} depends on whether the fluctuations of the field occur within a grand canonical or canonical ensemble. 
(For the exactly solvable case of the one-dimensional Ising model, the behaviour of the force emerging within the two ensembles is analysed in Refs.\!\nocite{DR2022,DTR2023}\citenum{DR2022} and \citenum{DTR2023}.)
The grand canonical ensemble is actually the one discussed in all previous sections. In particular, for equal {\bc}s the force is attractive within that ensemble, while it is repulsive in the canonical case\cite{GVGD2016}. This is naturally realised as a consequence of the conserved dynamics.\\[-3mm]

\emph{Colloidal particles.---} Beyond the instances discussed above with fixed or slightly displacing boundaries, one might be interested in cases in which these surfaces move according to the very same effect of the {\ccf}  acting on them, as it occurs, e.g., for colloidal particles. 
Imposing {\bc} of the kind discussed in Sec.~\ref{sec:CP-U} on moving, non-planar boundaries turns out to be  challenging. Thus,  it is convenient to think of the particles as localised, compact parts of the medium with properties (e.g., compressibility or viscosity), which differ from the rest of the medium\cite{FGDT2013,RKK2017}. Within these parts, a symmetry-breaking field is applied instead of imposing 
{\bc}s.\cite{BDG2022,VFG2022,VG2022} 
This portion of the medium then moves, e.g., according to an over-damped Langevin dynamics, which takes into account the total force acting on each particle due to its interaction with the (remaining part) of the medium and which, in thermal equilibrium, satisfies the fluctuation-dissipation theorem. 
Rather generically,  even in the simplest case, in which the fluctuating field is Gaussian and the particle acts on the medium only as an external field, the effective dynamics of such a single particle turns out to be described by a  Langevin equation with a \emph{non-linear} memory kernel. This determines the effective, time-dependent friction and the correlations of its fluctuations,\cite{BDG2022} and it spans an increasingly wide time interval as the critical point of the medium is approached. It causes, for example, an algebraic --- instead of exponential --- long-time relaxation of the position of a harmonically trapped particle when initially displaced from the centre of the trap, independently of the dynamics of the {\op}  field corresponding to model A or B.\cite{VFG2022}

The simple representation of a colloidal particle as outlined above allows one to investigate\cite{VG2022} the evolution of the effective interaction between two particles when they interact with the same Gaussian field. This is the case for the experiments discussed further below, in which two optically trapped colloids are in contact with a near-critical medium. 
In particular, one can consider the case that the position of one of the two particles is periodically modulated in time with, say, a harmonic function. This might induce (depending on the type of dynamics and whether the medium is critical or not) a non-harmonic motion of the other particle, which is kept away from the first one by a quadratic confining potential. At late times the field-mediated interaction synchronises  the motion of these two particles, and the nonlinear response of the second particle can be determined as a function of the driving frequency of the first one\cite{VG2022}. The response can be compared with the adiabatic approximation 
which assumes an instantaneous relaxation of the {\op}  to its equilibrium configuration appropriate for the actual arrangement of the particles. This comparison highlights the effects of \emph{retardation} in the propagation of the force due to the slow critical dynamics of the medium.

Beyond the simplified models discussed above, the investigation of the evolution of  {\ccf}s, which is relevant for the critical dynamics of actual physical systems, has to rely on numerical calculations, even within the mean-field approximation. This is  due to the increasing complexity of the evolution equations for more realistic models. 
For example, in the case of binary liquid mixtures, in addition to the {\op} $\phi$ one has also to account for the velocity field $\bm{v}$ of the mixture. The dynamics, which captures the critical properties of $\phi$  and $\bm{v}$, corresponds to the so-called model H,\cite{HH77} which includes hydrodynamics. Within the mean-field approximation one considers the evolution equation without noise, in which $\phi$ is locally conserved, i.e., it obeys the continuity equation with the current $\bm{J}= \phi \bm{v} - D \bm{\nabla} \delta \HH/\delta\phi$, where $\HH$ is the Landau-Ginzburg effective free energy given in Eq.~\eqref{eq:LGWwB}. The boundary terms are modified as described further below. The local velocity $\bm{v}$ of the medium (which will eventually encompass also the particles according to the idea discussed above) obeys the Navier-Stokes equation $\rho( \partial_t
+\bm{v}\cdot\nabla)\bm{v} = {\bm{f}} -\bm{\nabla}\cdot{\stackrel{\leftrightarrow}{\mbox{\boldmath$\sigma$}}}$,
with $\stackrel{\leftrightarrow}{\mbox{$\bm\sigma$}} = p\!\!\stackrel{\leftrightarrow}
{\mbox{$\bm I$}}-\eta[(\nabla{\bm{v}})^\dagger+\nabla{\bm{v}}]$,
where $\stackrel{\leftrightarrow}{\mbox{$\bm I$}}$ is the unit tensor, $\rho$ the mass density of the liquid (which is assumed to be spatially constant), and $\eta$ its viscosity. The
hydrostatic pressure $p$ facilitates to satisfy the incompressibility condition
$\nabla\cdot {\bm{v}}=0$, and ${\bm{f}}({\bm{x}})$ 
is the force density ${\bm{f}} = -\phi\bm{\nabla}\delta \HH/\delta \phi$.
According to the approach outlined above, the presence of $N$ particles labeled by the index  $i\in \{1, \ldots, N\}$, at positions $\{\x_i\}$ in the medium is represented by a spatial modulation $\chi(\x-\x_i)$ of its local properties, where $\chi(\x)$ is a smooth function which vanishes for $|\x|$ larger than the particle radius while it rapidly attains 1 upon decreasing $|\x|$ below that  radius. This function determines the boundary contribution to $\HH$ in the form $\sum_{i=1}^N\int \rmd^dx [ c_i \chi(\x-\x_i) \phi^2(\x) - h_i |\bm{\nabla} \chi(\x-\x_i)|  \phi(\x)]$ where $c_i>0$ plays the role of the surface enhancement and $h_i$ that of the surface field. We note that this expression generalises  the boundary term appearing in Eq.~\eqref{eq:LGWwB}. In addition,  $\chi$ controls the spatially varying viscosity $\eta(\x) = \eta_l + (\eta_c-\eta_l)\sum_{i=1}^N\chi(\x-\x_i)$ where
$\eta_l$ is the viscosity of the $l$iquid and $\eta_c\gg\eta_l$ the (common) viscosity of the $c$olloidal particles.
 The particles are assumed to be a region of space filled by a highly viscous fluid. (From this approach the notion of ``fluid particle dynamics'' earns its name; see, e.g., Ref.~\citenum{FGDT2013}). 
Similarly, the diffusivity $D$ acquires a space dependence, $D(\x) = D[1-\sum_{i=1}^N\chi(\x-\x_i)]$, which suppresses diffusion within the particles, while ${\bm{f}}$ (see above) receives a contribution $- \mathscr{V}^{-1}\sum_{i=1}^N\chi(\x-\x_i) \bm{\nabla}_{\x_i} \HH$, where $\mathscr{V} = \int \rmd^dx\, \chi(\x)$ is the effective volume of each particle. The resulting evolution equations for $\phi(\x,t)$ and $\bm{v}(\x,t)$ are solved numerically\cite{FGDT2013} together with the dynamics of the centre $\x_i$ of the particle $i$, which is determined by its instantaneous velocity $\bm{V}_i(t) =  \int \rmd^d x\, \bm{v}(\x,t) \chi(\x-\x_i)/\mathscr{V}$. 
In addition to the terms discussed above, a mechanical potential $U(\x_1,\ldots,\x_N)$  might be introduced in $\HH$ in order to account for possible direct interactions among the particles.

Based on the approach described above,  the authors of Ref.~\citenum{FGDT2013} investigated the temperature dependence of the drag force on a single particle steadily moving in a near-critical fluid mixture. They studied also the relative motion of two particles due to the non-equilibrium {\ccf} acting on them.
Due to the slow dynamics of the {\op} and depending on the P\'eclet number of the particle motion, significant deviations of the instantaneous force acting on the colloids from its equilibrium value at the instantaneous separation between the two particles are observed. These retardation effects reduce the strength of the force during particle aggregation. 

Before concluding this part, we mention that various simplifications of the model H described above, which are related to the velocity field $\bm{v}$, were considered in the literature. They allowed one to determine in semi-analytic form, e.g., the drag coefficient of a solid spherical colloid in the near-critical binary liquid mixture (see Ref.\!\nocite{YF2020}\citenum{YF2020} and references therein) or the amplitude of the fluctuations of the position of such a colloid when it is confined by an harmonic trap (see Ref.\!\nocite{FY2016}\citenum{FY2016} and references therein). 
In the latter case, it would be rewarding to investigate how such an amplitude emerges from the equilibrium distribution of the particle displacement, which is expected to be determined solely by the total potential of the mechanical forces acting on the particle.

\subsection{Experiments}
\label{ssec:Dyn-exp}

The dynamical aspects of {\ccf}s are experimentally still largely unexplored. This is primarily due to the fact that the relaxation times $\tau$ within experimental reach are still rather short compared to the time scales at which the experimental measurements are performed. Accordingly, as we shall discuss below, the available experimental evidences can still be interpreted in terms of the occurrence of equilibrium {\ccf}s, at least within the currently available experimental accuracy. No  retardation effects have been observed so far. However, the possibility of using these forces also for controlling the dynamics of colloidal particles was demonstrated.\cite{MDPC2017,MC-2019}

%
%
%
\begin{figure*}[h!]
\centering
  \includegraphics[width=0.8\textwidth]{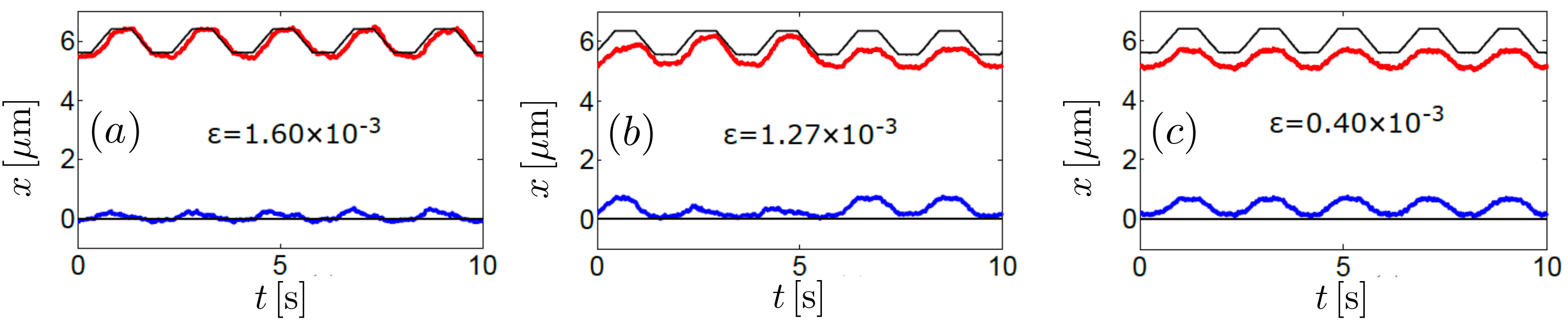}
 \caption{ %
Synchronisation of the motion of two trapped, identical colloidal particles (with radii $R\approx 2.5\mu{\rm m}$) upon approaching, from left to right, the (lower) critical temperature $T_c \approx 30^\circ{\rm C}$ of the solvent, constituted by an aqueous, critical micellar solution of the nonionic surfactant $\mbox{C}_{12}\mbox{E}_{5}$. The solid black lines indicate the  respective positions of the centres of the optical traps which confine the two silica colloids. The red and blue lines, instead, correspond to the experimentally determined centres of these two particles, while $\varepsilon = (T_c-T)/T_c$ quantifies the deviation from criticality.  Upon approaching criticality from left to right, the two particles increasingly displace from the centres of the corresponding optical traps under the action of the attractive {\ccf}s, reducing the inter-particle distance. This implies that the red line is systematically below the upper black curve while the blue line above the lower black one. Correspondingly, the motion of the particle, which is in that trap fixed at $x=0$, is increasingly influenced and synchronised with the motion of the  particle in the moving trap. (Figure adapted from Ref.~\citenum{MDPC2017}.)}
\label{fig:syn}
\end{figure*}
%

Critical Casimir forces can be used to synchronise the motion of two identical colloidal particles which are immersed in a binary liquid mixture and are trapped by two optical tweezers separated by a certain time-dependent distance. Upon moving periodically the centre of one of the two traps and therefore by moving one of the two colloids, it was shown\cite{MDPC2017} --- see Fig.~\ref{fig:syn} --- that it is possible to induce motion of the other colloid, in a way which depends on the thermodynamic distance $\varepsilon$ from criticality. If the mixture is far from being critical and thus the correlation length $\xi$ is smaller than the typical inter-particle distance $\ell$, the motion of one colloid does not actually affect the other, as expected and shown in Fig.~\ref{fig:syn}$(a)$. This also demonstrates that the hydrodynamic interactions are  negligible in this setting.\cite{MDPC2017} The situation changes when, upon approaching criticality, $\xi$ becomes comparable to $\ell$ and one colloid senses the presence of the other in terms of the action of {\ccf}s, as shown in panels $(b)$ and $(c)$ of Fig.~\ref{fig:syn}. Eventually, their motion synchronises and the average $\ell$ decreases because of the occurrence of these attractive forces.
This setting and the periodic protocol allows one to study experimentally also the energetics of the system composed of the two trapped colloidal particles and therefore to determine the distribution of work done on this system with the heat dissipated by each particle, which depends on the deviation from criticality.\cite{MPC2023} In particular, the temperature variations are obtained in Ref.~\citenum{MPC2023} by inducing local heating via a suitable laser. These relatively small variations produce large effects due to the extreme sensitivity of {\ccf}s on temperature.

Alternatively, {\ccf}s can also be used to influence the dynamics of diffusing colloidal particles in the absence of optical 
trapping.\cite{MC-2019} In particular, by using blinking optical tweezers, which can be periodically switched on and off, the dynamics of two colloidal particles in a near-critical binary liquid mixture was experimentally observed during diffusion. This way one can highlight the effect of their mutual interaction, possibly due to {\ccf}s. By studying the position of the particles when both traps are switched on, one can determine the non-universal parameters (e.g., trap stiffnesses, Debye screening length, critical temperature, etc.), which control, inter alia, the colloidal interaction in thermal equilibrium and which therefore fix its form. This interaction potential can then be used to predict, via numerical (Langevin or {\mc}) simulations, the statistical properties of the dynamics of the particles in the absence of the trapping. Then these predictions can  be compared with the actual experimental data for the relative velocity of the two particles in order to highlight the possible emergence of retardation effects.  However, as anticipated above, no significant discrepancy was found upon this comparison. A similar conclusion was reached on the basis of the data for the distribution of the first passage time, i.e., of the first time at which the inter-particle distance reaches a certain threshold value when they are released from an initial distance larger or smaller than this threshold. Depending on the temperature, one clearly observes that the emergence of the (attractive) critical Casimir interaction between the particles influences the distribution of these first passage times.\cite{MC-2019}
%


\section{Perspectives}
\label{sec:persp}

In the previous sections we focused on Casimir-like effects induced by the confinement of the critical fluctuations, which characterise a continuous phase transition occurring in a liquid medium at thermal equilibrium. Before concluding, we mention analogous instances in which Casimir-like forces emerge because of the fluctuations of other relevant collective fields. 
These include, for example, fluctuations in liquid crystals (see, e.g., Refs.\!\nocite{ZPZ98}\citenum{ZPZ98} and \!\nocite{KPHD2006}\citenum{KPHD2006} and references therein), in polymers\cite{U2001}, of the heights of interfaces\cite{ODD2005,LOD2006} and membranes\cite{BDF2010,BRF2011,MVS2012,BPK2017}, in complex fluids,\cite{GZS2013} 
and in granular materials\cite{CBMNS2006,BSM2007}, as well as fluctuations associated with non-equilibrium phase transitions\cite{BMS2007} or with transitions to non-equilibrium steady states\cite{KOS2013,AKK2015,KOS2015}.
More recently, motivated by the increasing interest in active matter, also this subject has been studied in the present context\cite{RRR2014,FGG2021}.  In particular, an early numerical study of an ensemble of run-and-tumble particles in $d=2$, confined by parallel lines of \emph{finite} length, revealed the emergence of an attractive force, the magnitude of which increases upon increasing the run length of the particles, but decreases \emph{exponentially} upon increasing the distance $L$ between the lines. This attraction may be turned into a repulsion via varying $L$ and the line length. 
In order to highlight the effects associated with confining an emergent collective behaviour within these systems, the phenomenon of flocking in active matter has been considered\cite{FGG2021} in $d=2$. In fact, a system of interacting self-propelled particles is known to undergo a flocking transition, characterised by the emergence of a well-defined flocking direction of collective motion (see Ref.\!\nocite{FGG2021}\citenum{FGG2021} and references therein). In the presence of two reflecting or partly repelling parallel walls, the flocking direction is parallel to them.  An attractive Casimir-like effect emerges on the walls, which decays \emph{algebraically} upon increasing their separation $L$, while displaying a certain degree of universality. However, a systematic understanding of the various features of such forces, caused by the long-ranged correlations of the fluctuations in active matter systems which thus are similar to those at criticality, is still lacking. Therefore this stands as an interesting problem for future investigations. 

In this review we discussed the emergence of {\ccf}s in soft matter, starting from very basic theoretical ideas up to reporting the first experimental evidences in wetting films, and, most importantly, in colloidal suspensions. 
These studies brought to light the following peculiar and useful features of {\ccf}s: (i) they can be reversibly switched on and off by controlling the temperature of the solvent; (ii) their strength and spatial direction can be controlled by suitable treatments of the \emph{surfaces} of the involved bodies;  (iii) they exhibit a universal character, and therefore they are largely independent of the actual material properties of the involved bodies. These features are rather unique to the {\ccf}s, especially when compared with other forces acting in soft matter, such as electrostatic, depletion, van der Waals, hydrodynamics forces etc. Most of them sensitively depend on the microscopic properties of the involved materials, and they can hardly be tuned, as far as their strength, range, and spatial orientation are concerned, by varying an external parameter. In this respect, {\ccf}s may find a number of possible applications  in  controlling and manipulating soft matter. In fact, this is the scientific direction into which most of the corresponding research has been developing recently. 
For example, the various studies mentioned in Sec.~\ref{ssec:FtoC-mb} focussed on exploiting {\ccf}s in order to control the assembly of colloidal dispersions and the possible structures of aggregates emerging from the assembly, with an eye on concrete, technical applications. 
In addition, a recent experimental study\cite{GSL2018} demonstrated that these forces provide a generic method for size-selective nanoparticle purification/separation. The aggregation they cause in a heterogeneous colloidal dispersion depends on the relative particle sizes. The larger colloids aggregate while the smaller ones remaining in solution. 
This allowed the authors to propose a novel approach which turned out to be very efficient in purifying commercial silica nanoparticles in a water-lutidine mixture. 
These investigations open the door to a number of possible applications of {\ccf}s to nanotechnologies.  
In this respect, it is also of high importance to understand how these forces can be used to counteract those which are usually at play in electromechanical devices at the micrometer or nanometer scale, i.e., MEMS or NEMS, respectively.  Among the latter forces, the most prominent  is the attractive Casimir-Lifshitz force, usually responsible for the undesired stiction of metallic parts. While in the experiments with colloidal particles described in Secs.~\ref{sec:exp} and \ref{ssec:Dyn-exp}, such a van-der-Waals force turned out to be negligible, this might not always be the case. 
This possibility motivated the experimental study presented in Ref.\!\nocite{CvC-2022}\citenum{CvC-2022}.
It involves a microscopic gold flake suspended above a flat gold-coated substrate and immersed in a critical water-lutidine binary liquid mixture. In turns out that the attractive Casimir-Lifshitz force acting on the flake --- which, on its own, cannot be easily made repulsive --- can be conveniently counterbalanced by 
the repulsive critical Casimir force. This offers an active control of the total force exerted on the flake. 
Beyond the cases discussed here and in Secs.~\ref{ssec:FtoC-el} and \ref{ssec:exp-salt}, the investigation of the interplay of {\ccf}s with other interactions in devices and in soft matter has still to reach a satisfactory level of understanding and quantitative accuracy. This stands as another interesting open problem for future studies. 

For example, it has been realised only recently\cite{SRP2022} that in previous theoretical studies of the wetting transition in systems with short-ranged interactions, a low-temperature Casimir contribution had been overlooked in the derivation of the interface potential describing the effective interaction between the free interface of the wetting film and a substrate. This contribution leads to a number of qualitative and quantitative changes compared to long-standing predictions. This resolves also a controversy concerning the comparison between corresponding theoretical predictions and available numerical simulations.   

In addition to some of the aforementioned open problems and directions for future research as discussed in Sec.~\ref{sec:Dyn}, the dynamics of {\ccf}s in and out of thermal equilibrium pose a theoretical and experimental challenge and are a largely unexplored subject beyond simplistic models. Accordingly, any progress in this direction, e.g., by accounting for the dynamics of fluctuations beyond mean-field theory or by considering more realistic geometries for their confinement, would be highly valuable, also in view of promising experimental access to these phenomena.\\[-1.5mm]

{\sl The scientific activity reviewed in this contribution was largely motivated by the inspiring work of Michael E.~Fisher, and it enjoyed his support and encouragement in the course of time. The wealth and the variety of theoretical and experimental results, which originated from these studies, are therefore part of his scientific legacy and 
pay tribute to his outstanding scientific oeuvre.}

\section*{Acknowledgements}
AG acknowledges support from MIUR PRIN project ``Coarse-grained description for non-equilibrium systems
and transport phenomena (CO-NEST)'' n.~201798CZL.

\balance

\providecommand*{\mcitethebibliography}{\thebibliography}
\csname @ifundefined\endcsname{endmcitethebibliography}
{\let\endmcitethebibliography\endthebibliography}{}

\end{document}